\documentclass{aastex}
\usepackage{emulateapj5}
\usepackage{apjfonts}
\usepackage{psfig}

\def\etal{{ et al.\thinspace}}

\def\gtrsim{\mathrel{\raise0.35ex\hbox{$\scriptstyle >$}\kern-0.6em
\lower0.40ex\hbox{{$\scriptstyle \sim$}}}}
\def\lesssim{\mathrel{\raise0.35ex\hbox{$\scriptstyle <$}\kern-0.6em
\lower0.40ex\hbox{{$\scriptstyle \sim$}}}}
\def\Msun{\hbox{$\rm\thinspace M_{\odot}$}}

\def\h50{h_{50}}

\def\ha{H$\alpha$}
\def\oii{[O{\sc ii}]}
\def\nii{[N{\sc ii}]}

\begin{document} 

\title{Galaxy Star-Formation as a Function of Environment in the Early Data Release of the Sloan Digital Sky Survey$^1$}

\shortauthors{G\'{o}mez \etal}
\shorttitle{Environmental Dependent Star Formation in the SDSS}

\author {Percy~L.~G\'{o}mez,$\!$\altaffilmark{2}
Robert~C.~Nichol,$\!$\altaffilmark{2}
Christopher~J.~Miller,$\!$\altaffilmark{2}
Michael~L.~Balogh,$\!$\altaffilmark{4} 
Tomotsugu~Goto,$\!$\altaffilmark{2,3}
Ann~I.~Zabludoff,$\!$\altaffilmark{5} 
A.~Kathy~Romer,$\!$\altaffilmark{2} 
Mariangela Bernardi,$\!$\altaffilmark{2,6} Ravi Sheth,$\!$\altaffilmark{6}
Andrew~M.~Hopkins,$\!$\altaffilmark{6}
Francisco~J.~Castander,$\!$\altaffilmark{7}
Andrew~J.~Connolly,$\!$\altaffilmark{6}
Donald~P.~Schneider,$\!$\altaffilmark{8} 
Jon Brinkmann,$\!$\altaffilmark{9} 
Don~Q. Lamb,$\!$\altaffilmark{10}
Mark SubbaRao,$\!$\altaffilmark{10}
Donald~G. York$\!$\altaffilmark{10}
\\ }

\altaffiltext{1}{Based on observations obtained with the Sloan Digital Sky Survey}

\altaffiltext{2}{Department of Physics, Carnegie Mellon University, 5000 Forbes Avenue, Pittsburgh PA 15213} 
\altaffiltext{3}{Institute for Cosmic Ray Research, University of Tokyo, Kashiwanoha, Kashiwa, Chiba 277-0882, Japan}
\altaffiltext{4}{Department of Physics, University of Durham, South Road, Durham DH1 3LE, UK}
\altaffiltext{5}{Steward Observatory, University of Arizona, 933 North Cherry Avenue, Tucson, AZ 85721}
\altaffiltext{6}{Department of Physics and Astronomy, University of Pittsburgh, Pittsburgh PA 14217}
\altaffiltext{7}{Institut d'Estudis Espacials de Catalunya/CSIC, Gran Capita 2-4, 08034 Barcelona, Spain}
\altaffiltext{8}{Department of Astronomy and Astrophysics, The Pennsylvania State University, University Park, PA 16802}
\altaffiltext{9}{Apache Point Observatory, 2001 Apache Point Road, P.O. Box 59, Sunspot, NM 88349}
\altaffiltext{10}{Astronomy and Astrophysics Department, University of Chicago, 5640 S. Ellis Ave, Chicago, IL 60637}

\setcounter{footnote}{10}

\begin{abstract}
We present in this paper a detailed analysis of the effect of environment on
the star--formation activity of galaxies within the Early Data Release (EDR)
of the Sloan Digital Sky Survey (SDSS).  We have used the \ha\ emission line
to derive the star--formation rate (SFR) for each galaxy within a
volume-limited sample of 8598 galaxies with $0.05\le z\le 0.095$ and $M(r^*)
\leq -20.45$.  We find that the SFR of galaxies is strongly correlated with
the local (projected) galaxy density and thus we present here the {\it
density--SFR relation} that is analogous to the density--morphology
relation. The effect of density on the SFR of galaxies is seen in three
ways. First, the overall distribution of SFRs is shifted to lower values in
dense environments compared with the field population.  Second, the effect is
most noticeable for the strongly star--forming galaxies (\ha\ EW $>5$\AA) in
the 75$^{th}$ percentile of the SFR distribution.  Third, there is a ``break''
(or characteristic density) in the density--SFR relation at a local galaxy
density of $\sim 1\,h^{-2}_{75}\,{\rm Mpc^{-2}}$. To understand this break
further, we have studied the SFR of galaxies as a function of clustercentric
radius from $17$ clusters and groups objectively selected from the SDSS EDR
data.  The distribution of SFRs of cluster galaxies begins to change, compared
with the field population, at a clustercentric radius of $3$--$4$ virial radii
(at the $>1\sigma$ statistical significance), which is consistent with the
characteristic break in density that we observe in the density--SFR
relation. This effect with clustercentric radius is again most noticeable for
the most strongly star--forming galaxies.

Our tests suggest that the density-morphology relation alone is unlikely to
explain the density--SFR relation we observe. For example, we have used the
(inverse) concentration index of SDSS galaxies to classify late--type galaxies
and show that the distribution of the star--forming (EW \ha\ $>5$\AA)
late--type galaxies is different in dense regions (within 2 virial radii)
compared with similar galaxies in the field.  However, at present, we are
unable to make definitive statements about the independence of the
density--morphology and density--SFR relation.

We have tested our work against potential systematic uncertainties including
stellar absorption, reddening, SDSS survey strategy, SDSS analysis pipelines
and aperture bias. Our observations are in qualitative agreement with recent
simulations of hierarchical galaxy formation that predict a decrease in the
SFR of galaxies within the virial radius. Our results are in agreement with
recent 2dF Galaxy Redshift Survey results as well as consistent with previous
observations of a decrease in the SFR of galaxies in the cores of distant
clusters. Taken all together, these works demonstrate that the decrease in SFR
of galaxies in dense environments is a universal phenomenon over a wide range
in density (from $0.08$ to $10\,h^{-2}_{75}\,{\rm Mpc^{-2}}$) and redshift
(out to $z\simeq 0.5$).
\end{abstract}

\section{Introduction} 

In this paper, we investigate the relation between environment and star
formation rate (SFR) of nearby galaxies.  Spectroscopic studies of distant
clusters of galaxies ($z>0.2$) have already established a clear connection
between these two physical properties, in that the SFR of galaxies in the
cores of distant clusters are significantly lower than those observed in the
field at the same redshift \citep{B+97,Hash98,P+99,C+01,PLO}.  In particular,
\citet{B+98} found that the SFRs of cluster galaxies were lower relative to
field galaxies of similar bulge-to-disk ratio, physical disk size, and
luminosity, which suggests that the observed decrease in the star--formation
may not be fully explained by the density--morphology \citep{Dressler} or
radius--morphology \citep{WGJ} relations (see also
\citet{P+99,C+01,A1689}). This conclusion has also been suggested by
\citet{Hash98}, who used the Las Campanas Redshift Survey (LCRS) to study the
relationship between local galaxy density, SFR, and galaxy concentration index,
and found that cluster galaxies have a reduced level of SFR compared with the
field, regardless of their concentration index. All these studies provide
circumstantial evidence that galaxies are undergoing a physical transformation
as they enter dense environments, and that the timescale for this
transformation is different for the morphological and star formation
properties.

In models of hierarchical galaxy formation \citep{KWG,SP99,Cole2000}, it is
assumed that galaxies accreted into larger halos ({\it e.g.,} they
fall into groups or clusters) have their hot gas reservoir removed, which
results in a gradual decline of the star--formation activity of galaxies in
dense environments.  This simple prescription is able to successfully match
some of the general observed trends \citep{semianal,Diaferio,infall}.  At
present, however, these models do not explore the physical details of how the
galaxies lose their gas (see Bekki, Shioya \& Couch 2002).  Many possible
mechanisms are proposed in the literature, such as ram pressure stripping of
the cold gas due to interaction with the intracluster medium \citep{GG,QMB},
galaxy harassment from high velocity encounters with other galaxies
\citep{harass}, tidal disruption \citep{BV} or galaxy evolution via mergers
and close encounters in in-falling poor groups \citep{Z+96,ZM98}.  Proper
treatment of these effects will likely affect the details of the predicted
relationships between galaxy properties and environment, but the relative
importance of the proposed evolutionary mechanisms can only be determined with
better data than has been available so far.

As a first step in addressing these issues, we use the Early Data Release
(EDR; Stoughton et al. 2002) of the Sloan Digital Sky Survey \citep[SDSS;
see][]{SDSS_tech_short} to further understand the relation between environment
and the star--formation rate of galaxies. The EDR is an excellent database for
such a study because it provides a large, homogeneous sample of galaxies, with
medium resolution, flux--calibrated spectra and 5 passband photometry, over
a wide range of environments. Such data allow for the computation of the
\ha\, and \oii\, equivalent widths (EW) and SFR for all galaxies in a
self--consistent way.  Our work expands upon earlier cluster studies, as
discussed above, in four key ways. First, we probe the rarefied field and poor
group regimes, which include the most common galactic environments in the
Universe. Second, we focus on the low redshift universe ($z \le 0.1$), which
provides a zero--point for the high redshift studies. Third, we are able to
follow the star--formation of galaxies continuously from the cores of rich
clusters and poor groups into the rarefied field. Finally, we can accurately
quantify the local galaxy density in a uniform manner that is not subject to
statistical background or luminosity corrections.  Our work is complementary
to earlier field studies using the 2dF Galaxy Redshift Survey \citep[2dFGRS,
see][]{2dF-sfr} and the LCRS \citep{Hash98} and explores the density--SFR
relation in greater detail than these previous works.

In Section \ref{sec-data}, we present a description of the SDSS EDR data, the
sample selection, and our SFR measurements. Our results are presented in
Section \ref{clusanalysis}. In Section \ref{discussion}, we discuss possible
systematic biases, the density--morphology relation, and models of
hierarchical structure formation. We also compare our work with previous
studies, {\it e.g.,} the 2dFGRS. In Section \ref{conclusions}, we present our
conclusions. In the Appendix, we discuss the details of selecting clusters and
groups from the SDSS data. We also consider the effects of systematic biases
that may influence our results, including the internal reddening of galaxies,
sample selection, data analysis pipelines, and luminosity and aperture biases.
Throughout this paper, we use ${\rm H_o} = 75\,{\rm km\, s^{-1}\, Mpc^{-1}}$,
$\Omega_{m}=0.3$ and $\Omega_{\Lambda}=0.7$ unless otherwise stated.

\section{Data}\label{sec-data}
\subsection{Early Data Release of the Sloan Digital Sky Survey}\label{EDR}
The Sloan Digital Sky Survey (http://www.sdss.org) is a 5 passband ($u^*$,
$g^*$, $r^*$, $i^*$, $z^*$) imaging and medium--resolution ($R\simeq 1800$)
spectroscopic survey of the Northern Galactic Hemisphere \citep[see][for
details]{SDSS_tech_short}. For technical details regarding the SDSS imaging
survey, the reader is referred to Gunn et al. (1998), Smith et al. (2002),
Pier et al. (2002) and Hogg et al. (2002). In June 2001, the SDSS publicly
released photometric and spectroscopic data comprise of nearly 50,000 spectra
of galaxies, stars and QSOs over 460 ${\rm deg^2}$ of sky. This release is
known as the Early Data Release (EDR) and is fully described in Stoughton et
al. (2002).

One of the unique features of the SDSS is the set of sophisticated data
analysis software pipelines \citep[see][Stoughton et al. 2002 for
details]{SDSS_software} used to reduce the raw images and spectra into large
catalogs of sources. Briefly, for each flux--calibrated SDSS spectrum, a
redshift is determined both from the absorption lines (via cross-correlation;
Heavens 1993) and the emission lines (via a wavelet--based peak--finding
algorithm; Frieman et al. in prep).  Once the redshift is known, the
spectroscopic pipeline estimates the continuum emission at each pixel using
the median value from a sliding box of 100 pixels ($\simeq 100$\AA) centered
on that pixel.  Emission and absorption lines are measured through the fitting
of a Gaussian, above the best--fit continuum, at the redshifted
rest--wavelength of the lines. In order to accommodate line blending, the SDSS
pipeline fits multiple Gaussians in the case of the \ha\ and \nii\ doublet and
the [O{\sc III}] doublet. Thus, for all the major emission/absorption lines in
the galaxy spectra, the spectroscopic pipeline provides an estimate of the
equivalent width (EW), the continuum level (at center of the line), a line
identification ({\it e.g.,} \ha\ ), a goodness--of--fit ($\chi^2$), and
the height and sigma of the fitted Gaussian (and the associated statistical
errors on all these quantities). These quantities are measured for all the
major stellar emission and absorption lines regardless of their formal
detection. In the Appendix, we present preliminary quality assurance tests of
these emission line measurements which show that the \ha\ and \oii\ fluxes and
EWs are robust.

When converting magnitude to flux, we have ignored the {\it asinh} magnitudes
of \citet{LGS} and treated the SDSS--EDR Petrosian magnitudes as traditional
AB magnitudes \citep{F+96}.  At the bright magnitudes used herein, the
difference between these two magnitude systems is less than 1\%, {\it i.e.},
less than the photometric accuracy of the SDSS (see Stoughton et al. 2002).

\subsection{Sample Selection}\label{selectioncriteria}
The reader is referred to \citet{strauss02} for a detailed description of the
spectroscopic target selection for the SDSS main galaxy survey. For the
analyses presented in this paper, we have used the EDR data discussed above
but with the following additional selection criteria. We begin with all
objects that have been spectroscopically confirmed as galaxies and have a
redshift confidence of $\ge0.7$ (see Heavens 1993). These criteria result in a
sample of 41,622 galaxies. We then reject 37 objects that have certain
redshift warning flags set in the database, {\it i.e.,} we exclude spectra
with the Z\_WARNING\_NO\_BLUE (no blue side of the spectrum) and
Z\_WARNING\_NO\_RED (no red side of the spectrum) flags (see Stoughton et
al. 2002). We also reject 1231 galaxies because they are duplicates.  We have
excluded another 37 galaxies that have $z^* > 22.83$, as these are likely
spurious detections in the SDSS (see Table 21 of Stoughton et al., 2002), and,
at this magnitude, the effect of the {\it asinh} magnitudes starts to become
noticeable.  This selection leaves us with 40,317 galaxies.  Finally, we
exclude 4014 galaxies for which the \ha\ emission line could not have been
measured and 3452 galaxies for which the \oii\ emission could not have been
measured. This is because these emission lines fall in masked regions of the
spectrum, due to cosmic ray hits, cosmetic defects in the CCDs or missing
data.

We define a volume-limited sample by further restricting the sample to
galaxies in the redshift range $0.05\le~z\le 0.095$ and more luminous than
$M(r^*) = -20.45$ (k--corrected, for ${\rm H_o} = 75\,{\rm km\, s^{-1}\,
Mpc^{-1}}$). This magnitude limit corresponds to $\simeq M^{\star}(r^*)+1$, assuming
$M^{\star}_{r^*}~=~-20.8+5\log{h}$ from \citet{Sloan_lf}.  The lower redshift
limit is imposed to minimize aperture bias (see the Appendix), while the upper
limit is the redshift where our luminosity limit equals the magnitude limit of
the SDSS ($r^*=17.7$; Strauss et al. 2002). These cuts leave us with a
volume--limited sample 8598 galaxies for our analyses and includes all types
of galaxies, not just emission line objects.

\subsection{Measuring the Star--Formation Rate in Galaxies}
\label{measurement}
The star formation rates of galaxies can be indirectly determined in various
ways, as reviewed by Kennicutt (1998a). The most successful methods rely on
correlations of the star formation rate with measurements of the far-infrared
luminosity (which arises from dust heated by star formation), the radio
luminosity (which results from synchroton emission associated with supernovae)
and indicators that are sensitive to the ionizing flux from massive stars. The
last category includes measurements of the ultraviolet continuum and the
fluxes of nebular emission lines.  Each of these techniques is subject to
different biases and calibration uncertainties, but does give consistent
estimates for ``normal'' galaxies when these effects are accounted for
\citep{CL_ext,Hopkins01,S+01}.  In this paper we focus on SFRs estimated from
the H$\alpha$ line. We also present the \oii\ EW measurements to
facilitate comparison with high redshift studies of galaxies.  Both lines are
sensitive to the metallicity and ionization levels of the gas, though the
problem is more significant for \oii.  Dust extinction is the largest source
of uncertainty, as both the effective optical depth and dust geometry must
be considered \citep{CL}.

\begin{figure}
\plotone{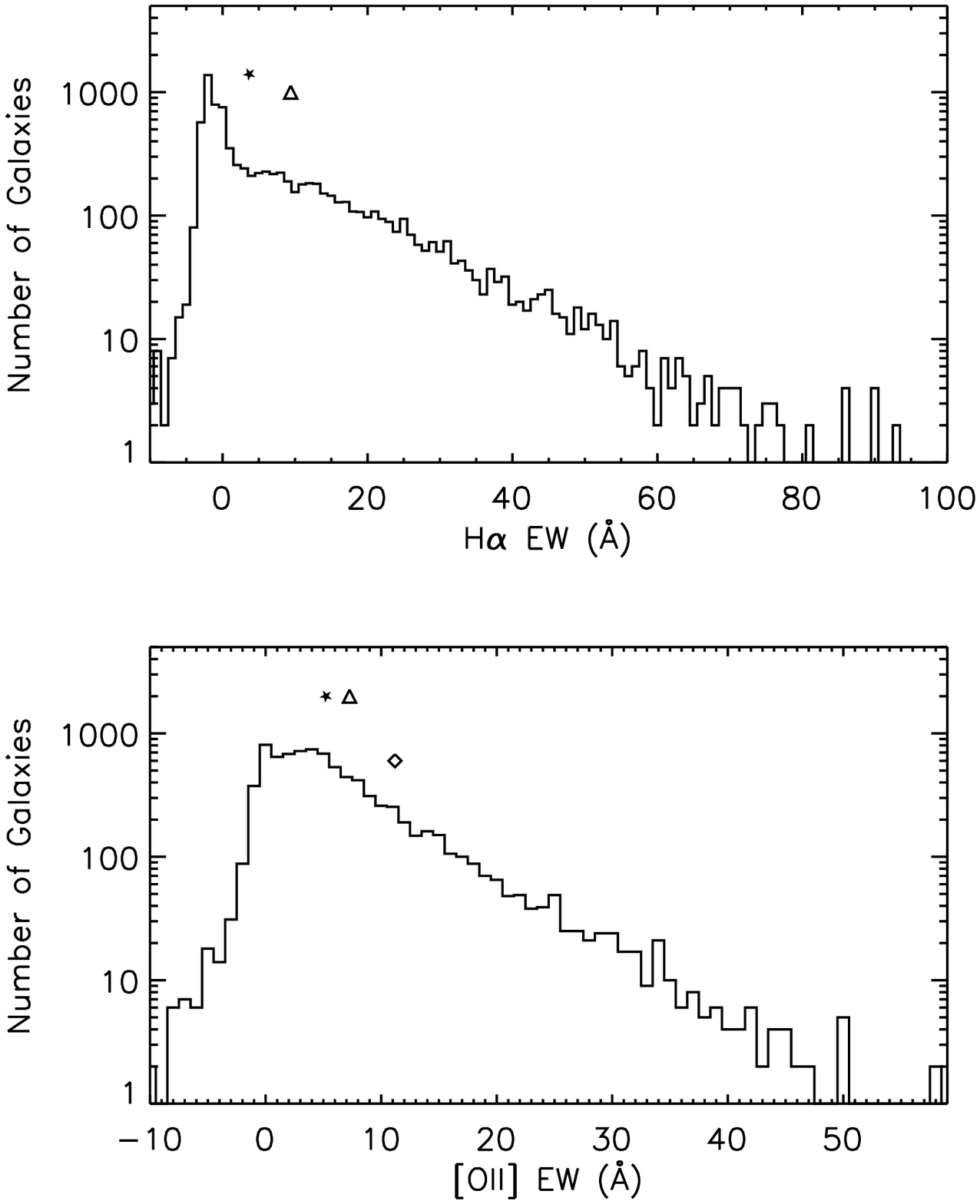}
\caption{ (Top) The distribution of observed \ha\ EW for the galaxies in our
sample. The star symbol is the median value, while the triangle symbol is the
mean.  The negative EWs in the distribution are caused by statistical
uncertainties and stellar absorption (see text for discussion).  (Bottom) The
distribution of observed \oii\ EW for the galaxies in our sample. The star
symbol is the median value, while the triangle symbol is the mean and the
diamond represents the mean field \oii\ EW from \citet{B+97}.  The negative
EWs in the \oii\, EW distribution are caused by statistical scatter.
\label{EWdist}
}
\end{figure}

In Figure \ref{EWdist}, we show the distribution of rest--frame \ha\ and \oii\
EWs, for the sample of galaxies discussed above.  There is a small tail of
negative EWs caused by statistical errors and, in the case of \ha\ , stellar
absorption. Typically, stellar absorption is expected to account for $<5$\AA\
of the \ha\ EW \citep{K92,CL_ext} and is therefore a small contribution to the
majority of the strong emission line galaxy spectra (see the Appendix for more
details).  The distributions of \oii\ and \ha\ EWs are highly asymmetrical,
skewed to low values but with a long tail to large, positive values.  The
observed mean of the SDSS \oii\ distribution (see Figure \ref{EWdist}) is half
the value of the observed mean for the field galaxy sample of the CNOC1
distant cluster studies \citep{B+97}. This difference may be due to the
differences in the apertures used by the SDSS and CNOC (although at z=0.1 the
SDSS aperture corresponds to 5.5 kpc, while at z=0.3 the CNOC aperture
corresponds to 6.7 kpc), or the evolution in the SFR of field galaxies
\citep{L96}.

We have derived SFRs for these galaxies using the \ha\ flux (as computed from
the pipeline output) and the theoretical relation from \citet{K98},

\begin{equation}
SFR (M_\odot /{\rm yr}) = 7.9 \times 10^{-42}\,\, {\rm L(H{\alpha})},
\label{kennicutt}
\end{equation}

\noindent where ${\rm L(H\alpha)}$ is the observed luminosity in the
\ha\ line (in erg ${\rm s^{-1}}$).  This relation is valid for
so-called Case B (optically thin) recombination, with a \citet{Sp}
initial mass function.  We make no correction for negative SFRs that
are caused by the small tail of negative \ha\ EWs in Figure
\ref{EWdist}.  Although nonphysical, the negative SFR of these
galaxies are included when computing the median, 25$^{th}$, and
75$^{th}$ percentiles of the SFR distributions because removing them,
or setting them to zero, would artificially skew the
distributions. Alternatively, we could add a constant amount of
stellar absorption to our sample but this would still leave a small
negative tail, due to statistical errors, and the exact amount of
stellar absorption to add remains unclear (see Appendix). We note here
that the SFRs presented in this paper are not corrected for the 3
arcsecond SDSS fiber aperture and are therefore, systematically lower,
by a factor of $\sim5$, compared to total SFRs derived from the radio
or by integrating the light from the whole galaxy (see Hopkins et al.,
in prep)

We then correct the SFR for SFR--dependent reddening using the empirical
formulae of \citet{Hopkins01}. We have also computed the normalized SFR
\citep[SFRN; see][]{PLO}, which is the observed SFR divided by the $z^*$
luminosity of each galaxy in our sample, {\it i.e.,} ${\rm SFRN =
SFR/L^{AB}_{z^*}}$ where ${\rm L^{AB}_{z^*}}$ is the k--corrected ($<0.2$
magnitudes) AB $z^*$--band luminosity (in ${\rm erg\,s^{-1}\,}$\AA$^{-1}$ at
the effective wavelength of the $z^*$ filter, $9049$\AA) computed via the
prescription outlined in \citet{F+96}. To make the units more intuitive, we
have re--normalized the $z^*$ luminosities by the characteristic $z^*$ AB
luminosity from \citet{Sloan_lf}, {\it i.e.,}
$L^{\star}(z^*)=6.34\times10^{41}\,{\rm erg\,s^{-1}\,}$\AA$^{-1}$.  As
discussed in \citet{PLO}, such near--IR luminosities are less affected by
recent star--formation and k--corrections and are more representative of the
underlying stellar mass of the galaxy (at a given redshift) than are
luminosities measured in bluer bands. The SFRN is therefore approximately the
SFR per unit stellar luminosity, which helps to remove any large luminosity
(or mass) bias from our results. We have also investigated the z-band
luminosity function (LF) as a function of galaxy density and find, for the
narrow range of bright z-band luminosities covered in this study, that the LF
remains unchanged (within the errors) as a function of local galaxy
density. This demonstrates that our results are unaffected by possible changes
in the luminosity function of galaxies with local galaxy density.

\section{Results}\label{clusanalysis}

In this section, we present the results of two analyses of the relation
between environment and the star--formation rate of galaxies. The first is an
investigation of how the distribution of galaxy SFRs changes as a function of
local (projected) density. This analysis spans over two orders of magnitude in
density, from the rarefied field ($0.08\,h^{-2}_{75}\,{\rm Mpc^{-2}}$) to the
densest environments ($10\,h^{-2}_{75}\,{\rm Mpc^{-2}}$).  The second analysis
focuses only on virialized systems with velocity dispersions ranging from
$\simeq 200\, {\rm km\,s^{-1}}$ to $ 1000\, {\rm km\,s^{-1}}$ and investigates
the SFR of member galaxies as a function of clustercentric radius.

\subsection{SFR as a Function of Local Galaxy Density}
\label{densityanalysis}

We investigate the SFR of galaxies as a function of their local galaxy
density.  We can parameterize the local environment of each galaxy using the
projected density of galaxies in a manner similar to that employed by
\citet{Dressler}. This involves calculating the projected distance to the
10$^{th}$ nearest spectroscopically--observed neighbor, and then converting
that into a surface density ($h_{75}^{-2}\,Mpc^{-2}$). As discussed in Section
\ref{selectioncriteria}, our limit is $M(r^*)=-20.45$ ($h=0.75$), which is
close to the limit used by Dressler ($M(V)=-20.4$, $h=0.5$).
Our methodology does differ from that used by Dressler (1980) in that we use a
circular search aperture (instead of a rectangle) and spectroscopically
eliminate foreground and background galaxies so no further background
correction is necessary.  Thus our tenth nearest neighbor will be farther away
than obtained via Dressler's definition.  Therefore, our effective
``smoothing'' scale is larger, and we may not be as sensitive as Dressler
(1980) to the densest environments, {\it i.e.,} cluster cores.

When calculating the local density, edge effects are particularly important
because the EDR geometry is long strips of data that, at the median redshift
of our sample, subtend only $14.8\,h_{75}^{-1}\,$ Mpc on the sky. To account
for this, we remove any galaxy in which less than 75\% of the circular area
including the $10^{th}$ nearest neighbor lies within the EDR boundary.  This
corrections removes 1972 galaxies, leaving us with 6626 galaxies for our
analysis.  We do not make an area correction to our density estimates because
we do not know if the $10^{th}$ nearest neighbor we have observed is, in fact,
the true $10^{th}$ nearest neighbor. If it is, then correcting the density
estimates for the area outside the survey boundary would bias the density
estimates of those galaxies. Instead, we accept the fact that $\sim25$\% of
galaxies near the edge of the survey have an incorrect local density
measurement but that they are not systematically biased. We have checked the
effect of changing our tolerance to accepting galaxies near the edge of the
survey, from rejecting all galaxies where the $10^{th}$ nearest neighbor is
closer than any survey boundary, to including all galaxies regardless of edge
effects, and find our results are robust.

\begin{figure}
\epsscale{0.7}
\plotone{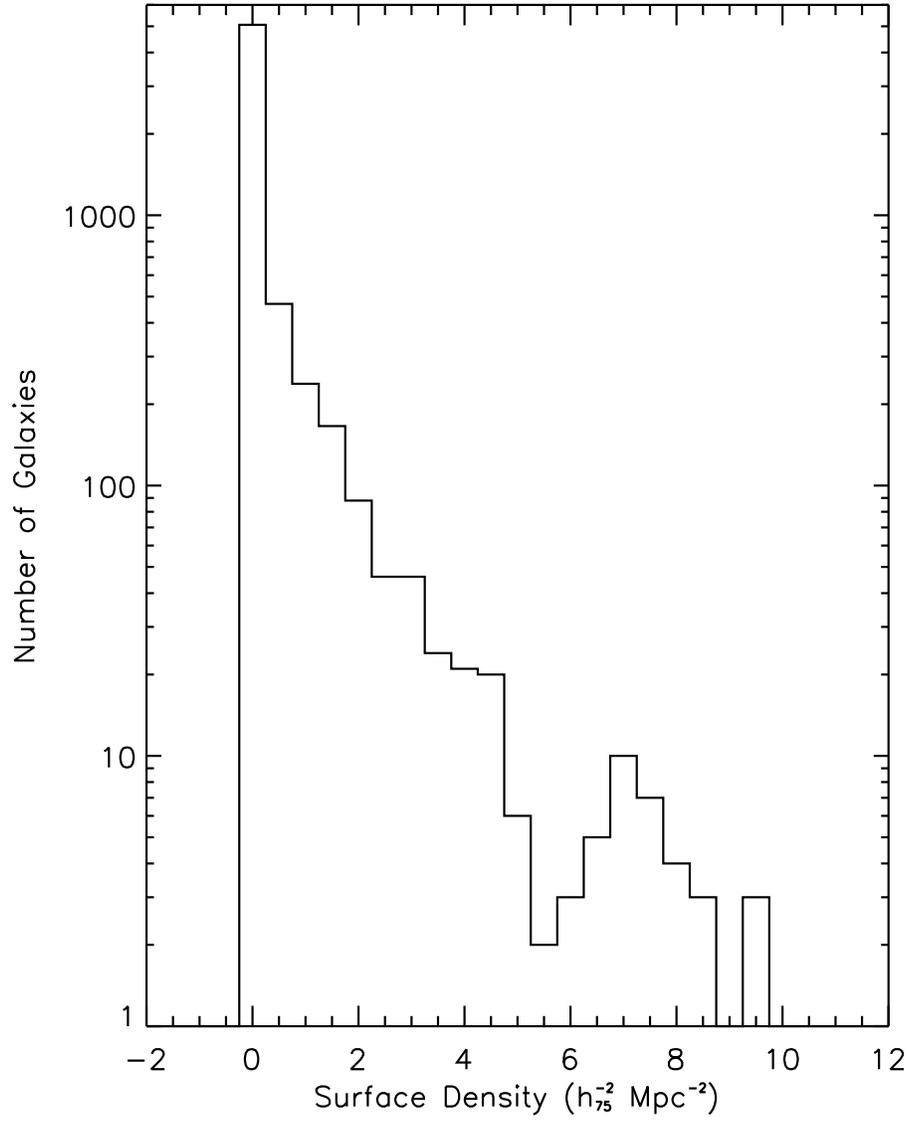}
\caption{The distribution of local galaxy densities ($h_{75}^{-2}\,Mpc^{-2}$)
for galaxies in our sample.
\label{densitydist}
}
\end{figure}

In Figure \ref{densitydist}, we show the distribution of local galaxy
densities for the 6626 galaxies that satisfy this constraint.  These local
galaxy density measurements are significantly lower than those presented by
Dressler (1980) and \citet{D+97}. These differences are partly due to the
cosmologies used, the definitions of the field and search area (see above),
differences in the edge corrections, and probable errors in the magnitudes used by Dressler (1980), which have uncertainties of at least 0.5
magnitudes. However, the most dominant effect is that we are probing
lower density regions than Dressler (1980).

\begin{figure}
\epsscale{0.8}
\plotone{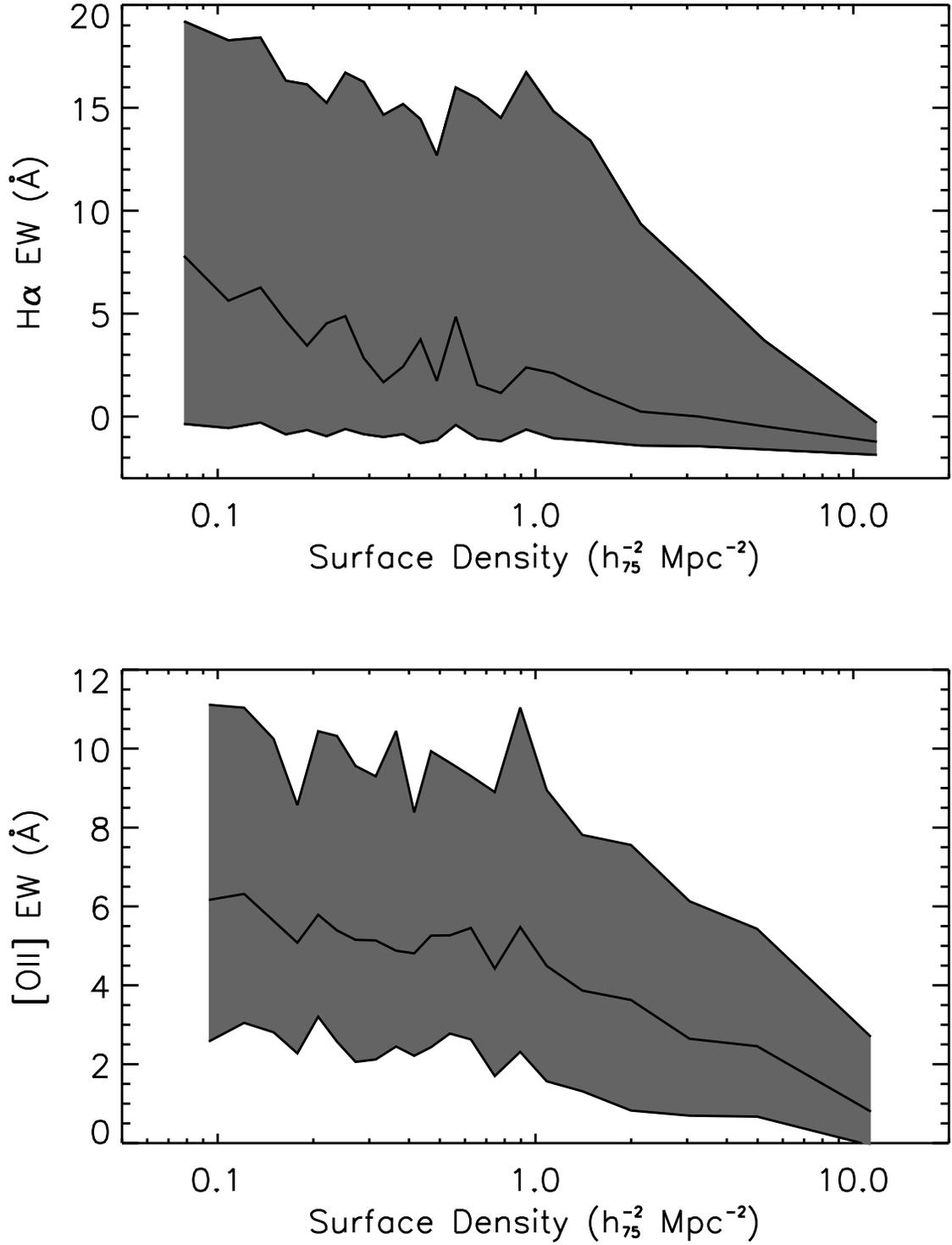}
\caption{(Top) The shaded area represents the distribution of \ha\ EW as a
function of the projected local surface density of galaxies. The top of the
shaded area is the 75\,$^{th}$ percentile of the \ha\ EW distribution, while
the bottom is the 25\,$^{th}$. The median is shown as a line.  We have used
all available galaxies in the SDSS EDR that satisfy our selection criteria
regardless of their location near a known cluster of galaxies.  We have,
however, excluded galaxies close to the survey boundary (see the text for a
complete explanation) and those that may have an AGN (see text). Each bin
contains 250 galaxies.  (Bottom) The same as above, but for the \oii\ emission
line. Each bin contains 150 galaxies.
\label{densityew}  
}
\end{figure}

In Figure \ref{densityew}, we present the \ha\ and \oii\ EW distributions, as
quantified by the median, the 25$^{th}$, and the 75$^{th}$ percentiles, as a
function of local density as defined above.  We see an overall shift of these
distributions to lower values in high density regions.  Furthermore, the most
strongly star--forming galaxies, {\it i.e.,} the tail of the distribution for
\ha\ EW $>5$\AA, appears to experience the largest effect. To understand this
effect further, we consider the (inverse) concentration index $C$ of SDSS
galaxies \citep{SDSS_morph}.  Although concentration will be sensitive to
non-morphological features like the presence of nuclear star-formation, there
appears to be a reasonable correlation between $C$ and the Hubble
morphological classifications as shown in Figure 10 of Shimasaku et
al. (2001). For the analysis in this paper, we have chosen a threshold of $C
>0.4$ to define late--type galaxies, instead of $C >0.33$ as proposed by
Shimasaku et al. (2001), as our choice provides a cleaner, but more
incomplete, sample of such galaxies. We estimate from Figure 10 of
\citet{SDSS_morph} that our higher threshold ensures less than 5\%
contamination from early--type galaxies in the late--type sample. We find that
the tail of the \ha\ EW distribution is dominated by late--type (spiral)
galaxies regardless of local density, {\it i.e.,} $> 75$\% of all galaxies
with \ha\ EW\, $>5$\AA\, are consistent with being late--type galaxies.

We note here that the $25^{th}$ percentiles of the \ha\ and \oii\ lines appear
to follow different slopes with density, with the \ha\ $25^{th}$ percentile
showing almost no change with density. This is because the \ha\ EWs at
$\lesssim 5$\AA\, are dominated by stellar absorption, and the strength of the
absorption is only weakly dependent on the stellar population for galaxies
older than a few Gyr.  On the other hand, the \oii\ EW is unaffected by
absorption and thus clearly shows the continual decline of emission flux into
the densest regions.

\begin{figure}
\epsscale{1.05}
\plottwo{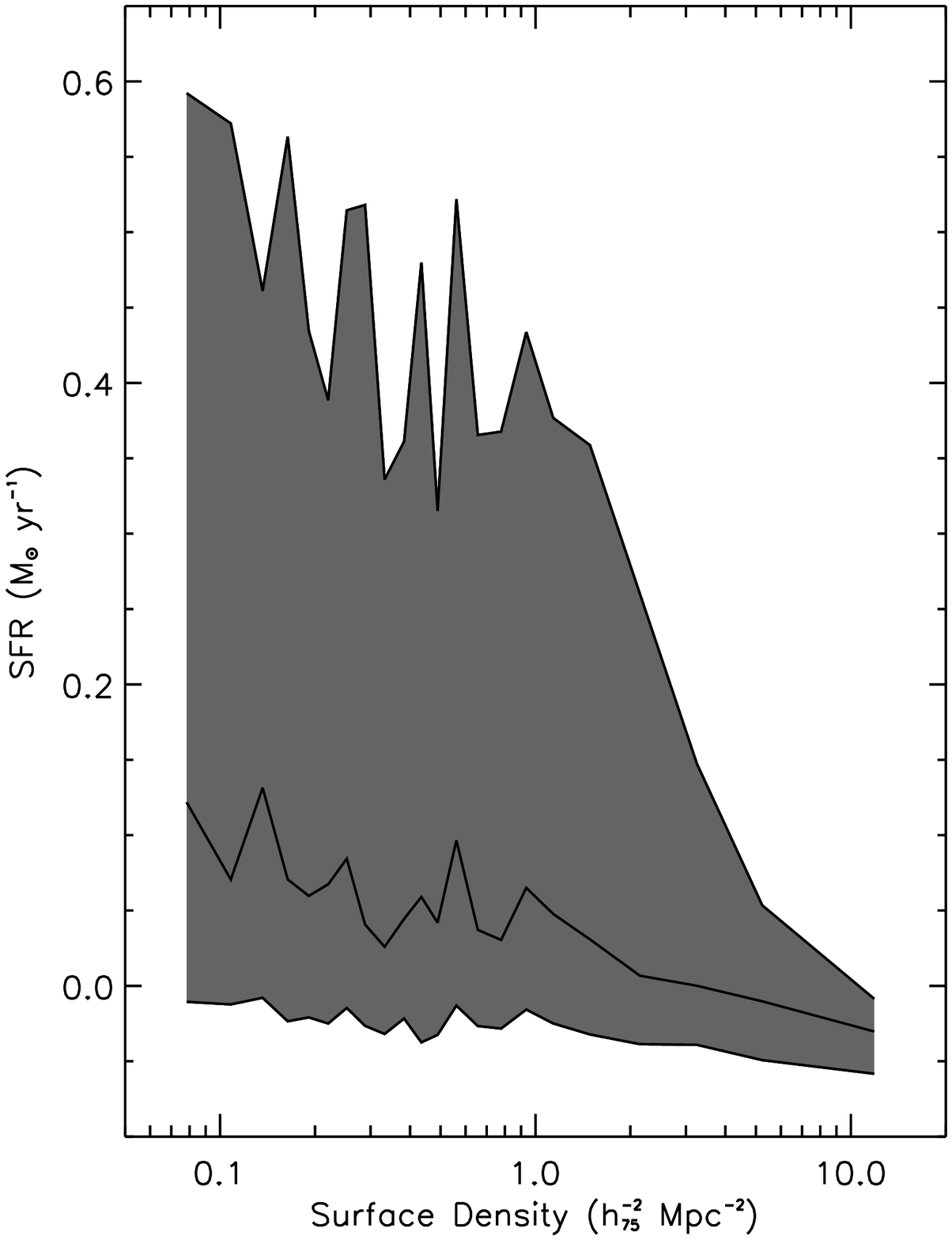}{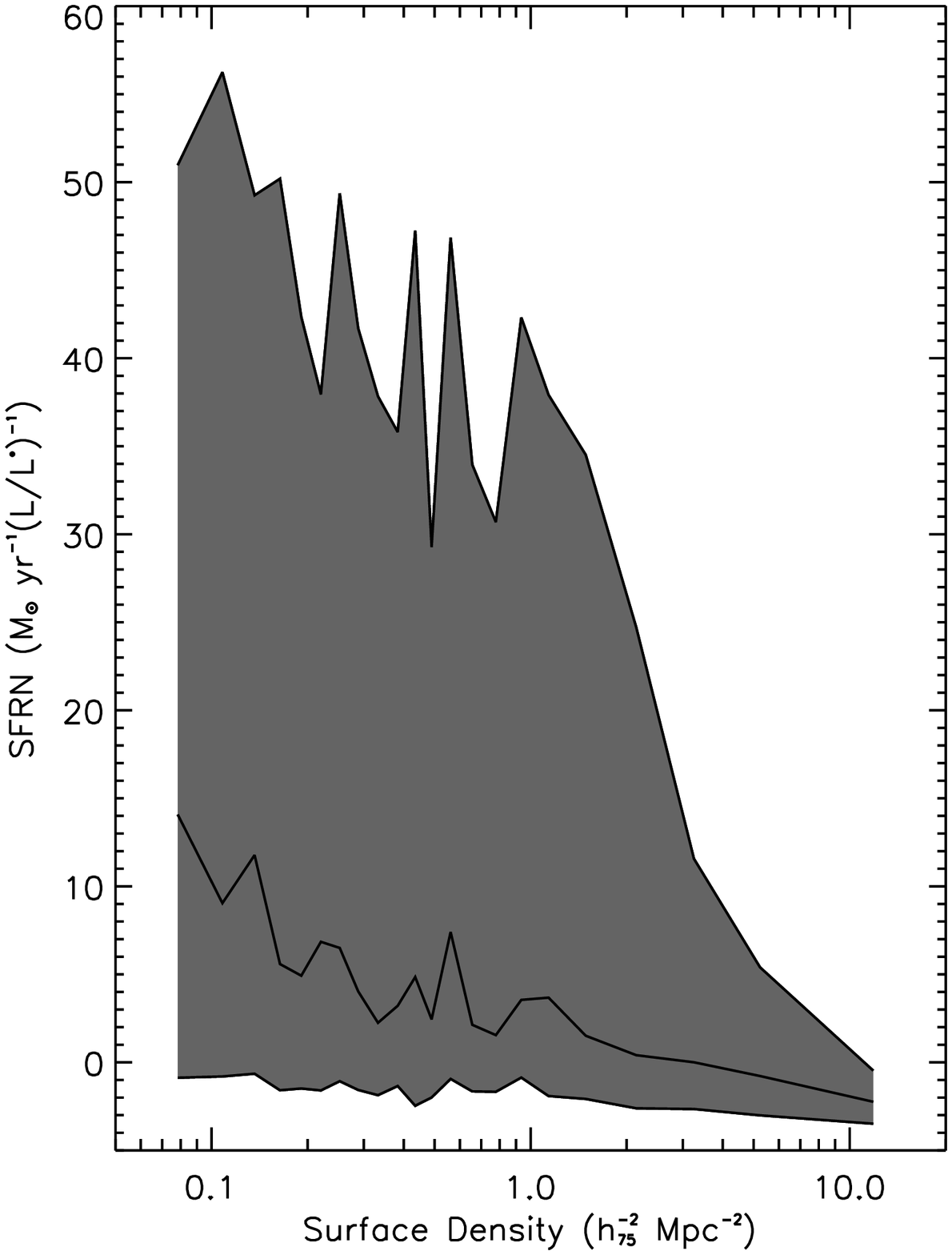}
\caption{(Left) The shaded area represents the distribution of
corrected SFR (Hopkins et al. 2001) as a function of the projected
local surface density of galaxies. (Right) The shaded area represents
the distribution of SFRN (the normalized SFR; see text) as a function
of the projected local surface density of galaxies. In both plots, the
top of the shaded area is the 75\,$^{th}$ percentile, while the bottom
is the 25\,$^{th}$ percentile. The median is shown as a solid line.
We have used all available galaxies in the SDSS EDR that satisfy our
selection criteria.  We have excluded galaxies near the edge of the
survey and those which may have an AGN present based on the Kewley et
al. (2001) prescription. Each bin contains 150 galaxies.  These plots
represents the density--SFR relation that is analogous to the
density-morphology relation of Dressler (1980). We note here that the
SFRs presented here are not corrected for the 3 arcsecond
SDSS fiber aperture and are therefore, systematically lower, by a
factor of $\sim5$, compared to total SFRs derived from the radio or by
integrating the light from the whole galaxy (see Hopkins et al., in
prep)
\label{densitysfr}
}
\end{figure}

In Figure \ref{densitysfr}, we show the density--SFR relation for galaxies
that is analogous to the density--morphology relation of Dressler (1980)
(although the density parameters are not directly comparable).  For this
analysis, we have also removed obvious Active Galactic Nucleii (AGN) using the
prescription outlined in Kewley et al. (2001) and using the [N{\sc ii}],
[O{\sc iii}], \ha\ and $\rm H_{\beta}$ emission lines.  If the detection of an
AGN was ambiguous, based on the prescription of Kewley et al. (2001), then we
do not exclude the galaxy from Figure \ref{densitysfr}.  We also show the SFRN
of galaxies as a function of local galaxy density, which shows two interesting
phenomena: First, when compared to the corrected SFR, there is a much larger
difference between the low and high density regions. This is likely an effect
of the lower luminosity galaxies in the field having a higher rate of SFR per
unit stellar luminosity than the more luminous galaxies. Secondly, the
``break'', or characteristic density, seen in the density--SFR relation at a
galaxy surface density of $\sim 1\,{\rm h_{75}^{-2}\,Mpc^{-2}}$ is much more
prominent. At densities lower than this ``break'', the SFR of galaxies
continues to increase, {\it i.e.}, in Figure \ref{densitysfr}, the median and
75$^{th}$ percentile of the SFRN distribution keeps increasing all the way to
the lowest densities studied in this paper ($0.08\,h^{-2}_{75}\,{\rm
Mpc^{-2}}$).

\subsection{SFR as a Function of Clustercentric Radius}
\label{radialanalysis}
We now turn our attention to virialized systems like clusters and groups of
galaxies, and investigate the SFR of galaxies as a function of clustercentric
radius. We test whether the features observed in the density--SFR relation,
({\it e.g.,} the break or characteristic density at $\sim 1\,{\rm
h_{75}^{-2}\, Mpc^{-2}}$) are also present in the radius--SFR relation around
known virialized systems.  The details of how we objectively select clusters
and groups from the SDSS EDR data are in the Appendix. We also discuss the $17$ clusters and groups used in this paper in the Appendix.

In Figure \ref{clusterew}, we present the observed distribution of \ha\ and
\oii\ EW, as quantified by the median, 25$^{th}$, and 75$^{th}$ percentiles, as
a function of (projected) clustercentric radius for $17$ groups and clusters
of galaxies detected in the EDR data (see Appendix).  To accommodate the wide
range of masses spanned by our cluster sample, we have re--scaled all radial
distances by the virial radius ($R_{v}$) of each cluster.  $R_{v}$ is computed
from the line--of--sight velocity dispersion, in units of km s$^{-1}$, using
the formula $R_{v}\simeq 0.002\sigma_r\,h_{100}^{-1}\,$ Mpc from
\citet{Girardi98}. For this calculation, we used $\sigma_r(1)$ as discussed in
the Appendix. For consistency with the higher redshift work, we provide
measurements for both the \oii\ and \ha\ emission lines.  In Figure
\ref{clustersfr}, we show the same as in Figure \ref{clusterew}, but now for
the SFR and SFRN of galaxies as a function of (projected) clustercentric
radius.

For comparison, we have also constructed a ``non--cluster'' (or field) sample
of galaxies that consists of all galaxies within our volume--limited sample
that are located, in redshift space, within $3.5\sigma_r$ of a cluster but are
at a (projected) clustercentric distance of $>25\,R_{v}$ from any of our
clusters.  This methodology guards against potential redshift selection
effects and/or redshift evolution of the field population, because we have
defined our field population in the same redshift shells as the cluster
galaxies.  We note however that the difference in the mean (or median) SFR of
galaxies of the field population as defined using our method and that of
taking all galaxies regardless of the presence of clusters, is less than 10\%
(with our mean SFR measurement being systematically higher as expected). Our
derived field values for the 25$^{th}$ and 75$^{th}$ percentiles are shown as
lines on Figures \ref{clusterew} and \ref{clustersfr}.

Figures \ref{clusterew} and \ref{clustersfr} show a clear decrease in the
star--formation activity (in either the SFR or the EW of \ha\ and \oii) as a
function of clustercentric radius.  As seen in Figures~\ref{densityew} and
\ref{densitysfr}, the SFR of galaxies in dense regions differ from that in the
field in two ways: the whole distribution of SFRs (and EWs) is shifted to
lower values, and the skewness of the distributions decreases, with the tail
of the distribution containing high SFR galaxies diminishing as one enters
denser environments. As discussed in Section \ref{densityanalysis}, this tail
of strongly star--forming galaxies (\ha\ EW $>5$\AA) is dominated by
late--type galaxies. For comparison with higher redshift studies, the mean
(median) of the \ha\ and \oii\ EW distributions within one virial radius of
the clusters studied herein are 3.4\AA\ ($-0.7$\AA) and 3.2\AA\ (1.8\AA),
respectively.

The results in Figures \ref{clusterew} and \ref{clustersfr} are
qualitatively similar to those for distant ($z\sim 0.3$) clusters of galaxies
\citep{B+97}.  A new aspect of our study, however, is that we can map
this decrease in SFR all the way from the cluster cores into the field
population.  We have used the Kolmogorov-Smirnov (KS) statistic to test the
distributions of EWs and SFRs, in each radial bin of Figures \ref{clusterew}
and \ref{clustersfr}, against the distribution of EWs and SFRs derived from our
field population. This test is designed to look for global differences between
two distributions and allows us to determine the clustercentric radius at
which the difference between the two distributions becomes statistically
significant. The distribution of \ha\ EWs becomes different from the
field at the 68\% ($>1\sigma$) level at $3.2R_{v}$, while for \oii\ the
distributions become different at $3.7R_{v}$.  For SFR and SFRN, the effect of
the cluster environment becomes noticeable, at the $>1\sigma$ level, at
clustercentric radii of $2.8R_{v}$ and $2.4R_{v}$, respectively.

It is evident from Figures \ref{clusterew} and \ref{clustersfr} that the tails
of these distributions are more strongly environment-dependent than the
medians. Therefore, we have performed a separate statistical test on the tails
of the distribution and how they change with clustercentric radius. We
calculated the difference between the 25$^{th}$ and 75$^{th}$ percentiles in
each radial bin of Figures \ref{clusterew} and \ref{clustersfr}, and the
corresponding 25$^{th}$ and 75$^{th}$ percentiles for the field (shown as
solid, straight lines in Figures \ref{clusterew} and \ref{clustersfr}.)  In
order to assess the statistical significance of these differences, we created,
for each radial bin, 1000 fake datasets via bootstrap re--sampling (with
replacement) and computed, for each dataset, the same difference between
cluster and field for both the 25$^{th}$ and 75$^{th}$ percentiles.  This
exercise provides a distribution of differences (between the cluster and
field) for each radial bin for both the 25$^{th}$ and 75$^{th}$ percentiles.
We then determined the probability that the observed percentile difference is
consistent with zero, {\it i.e.,} we just count the fraction of percentile
differences above zero in these fake datasets. Using this methodology, we
determined, at a greater than 68\% confidence ($>1\sigma$), that the 75$^{th}$
(25$^{th}$) percentile of the \ha\ EW distribution becomes statistically
different from the field population at a clustercentric radius of $3.5R_{v}$
($0.5R_{v}$). For the SFR distribution, the 75$^{th}$ (25$^{th}$) percentile
becomes statistically different from the field population at a clustercentric
radius of $3.6R_{v}$ ($0.5R_{v}$). For the SFRN, the 75$^{th}$ (25$^{th}$)
percentile becomes statistically different at a clustercentric radius of
$3.9R_{v}$ ($0.9R_{v}$)

These tests demonstrate that the SFR distribution of the galaxy population
differs from the field (at $1\sigma$) out to $\sim 3-4$ virial radii. These radii
correspond to $\sim 4\,h_{75}^{-1}\,$ Mpc, based on the average virial radii
computed for the clusters given in Table \ref{tabclus}. These
tests confirm that all levels of SFR are affected, with the most strongly
star--forming galaxies being affected more severely ({\it i.e.,} the 75$^{th}$
percentile) than the quiescent population. 
The implications of these results will be
discussed in Section \ref{discussion}.

To fully understand the correspondence between the radius--SFR relation and
the density--SFR relation presented in Section 3.1, we present the correlation
between the local projected density and clustercentric radius measured for
each galaxy within the vicinity of our 17 clusters and groups. As expected,
there is a linear correlation between these two quantities, and the
characteristic density of $\sim 1\,h_{75}^{-2}\,{\rm Mpc^{-2}}$ in Figure
\ref{densitysfr} corresponds to a range of clustercentric radii from $\sim2$
to $3$ virial radii. This range of radii is consistent with our observations
above in that the distribution of SFRs in cluster galaxies begin to differ
from the field population at $\simeq3$ to $4$ virial radii.


\section{Discussion}\label{discussion}

\subsection{Possible Biases and Systematic Errors}
Before we interpret our results, we must first investigate the effects of
possible systematic biases on our work. These include the effect of: {\it i)}
The SDSS analysis software and the SDSS survey observing strategy; {\it ii)}
Aperture bias due to the 3 arcsecond fiber width; {\it iii)} Reddening due to
dust in the host galaxies; and {\it iv)} The smearing correction performed by
the SDSS to improve the flux calibration of the continuum. We present a
detailed analysis of these possible biases in the Appendix, but we find
that none of these potential biases have a
large effect on our results.

\subsection{The Density--Morphology Relation}
\label{DMchap}

As we move towards denser environments, the galaxy population becomes
dominated by early-type galaxies that have a lower intrinsic SFR
\citep{K83,Jansen}.  In this section, we attempt to determine whether or not
the SFR of galaxies of a given morphology are themselves affected by
environment, {\it i.e.,} is the density--SFR relation distinct from the
density--morphology relation? Previously, \citet{B+98} have attempted to
answer this question using an objective morphological classification, based on
a galaxy bulge-to-disk luminosity, that was then correlated with the radial
dependence of SFR in distant clusters.  They concluded that cluster galaxies
of a given redshift, luminosity, and bulge-to-disk ratio had lower star
formation rates compared with their counterparts in the field.  Some evidence
supporting this has also been reported by \citet{Hash98}, \citet{C+01},
\citet{A1689} and \citet{kap2}.  We have attempted to address this issue in
this paper in two different ways. First, using a subset of SDSS galaxies that
possess both a visual morphological classification (from either Dressler 1980
or Shimasaku et al. 2001) and a SDSS \ha\ EW measurement. Secondly, using
objective morphologies based on the (inverse) concentration index of SDSS
galaxies as discussed in Section \ref{densityanalysis}. We discuss these two
tests below.

We have cross--correlated our sample of galaxies with samples of Dressler
(1980) and Shimasaku et al. (2001). For the \citet{Dressler} cluster sample,
we find 114 galaxies in common with our sample and thus have a SDSS \ha\
measurement.  The sample includes $23$ ellipticals, $39$ spirals and $50$
lenticulars that satisfy our selection criteria (Section
\ref{selectioncriteria}).  For the Shimasaku et al. (2001) sample, which
provides morphological classifications for 456 SDSS galaxies, selected over a
representative range of environments, we find 57 early--type galaxies and 73
late--type galaxies that satisfy our selection criteria (outlined in Section
\ref{selectioncriteria}). In this case, we have added together the ellipticals
and lenticulars because of the concerns noted by Shimasaku et al. (2001)
regarding their tendency to preferentially classify lenticular as ellipticals,
compared with the RC3 catalog. Overall, these two samples of
morphologically--classified galaxies cover a similar range in luminosity and
redshift as the galaxy sample outlined in Section \ref{selectioncriteria}.
The largest systematic uncertainty is in the consistency between the
morphological classifications.

In Figure \ref{morphs}, we show the distribution of \ha\ EWs for these galaxy
samples as a function of their morphological type.  We see a 
difference (confirmed using a KS test) in the \ha\ EW distribution of spirals
between these two samples, {\it i.e.,} the Dressler sample of spirals has, on
average, a lower SFR than seen in the Shimasaku et al. (2001) sample.  This
difference could be due to one or more of the following possibilities.
First, it could reflect a real environment effect, that cluster spirals have
lower SFR than the average field spiral population.  Second, it could be due to the
fact that the morphological bins are too coarse to define a homogeneous galaxy
population.  A different relative distribution of Sa and Sc galaxies in the
two samples, for example, could give rise to the same trend \citep[see][]{SJ}.
Third, the difference may reflect inconsistencies in the morphological
classification criteria in the two samples.  Thus, although these results are
intriguing, a more definitive answer will require an automated classification
of the SDSS galaxy morphologies, on a continuous scale.

As a first attempt of this, we have used the (inverse) concentration index
discussed in Section \ref{densityanalysis} to investigate the degeneracy
between the density--morphology and the density--SFR relations. This galaxy
parameter can be used as an objective morphological classification, but is
hard to interpret and to compare directly to the visual morphologies. In
Figure \ref{newfigure}, we present the distribution of SFRN as a function of
projected local galaxy density (as originally presented in Figure
\ref{densitysfr}), but split the sample into two broad morphological bins
based on the (inverse) concentration index, {\it i.e.}, late--type galaxies
(with $C>0.4$, see Section \ref{densityanalysis}) and early--type galaxies
(with $C\le0.4$). As discussed in Section \ref{densityanalysis}, the tail of
the SFR distribution (with high \ha\ EWs) is dominated by late--type galaxies
at all densities. This plot also demonstrates that the late--type galaxies in
our sample lie on a density--SFR relation similar to the one observed for the
whole sample. As expected, the early--type galaxies have little, or no,
star--formation, but they also appear to follow a shallow density--SFR
relation (for this sample however, we must be concerned about contamination
from the late--type galaxies and the effects of stellar absorption).

To quantify these effects in more detail, we present in Figure \ref{histo},
the distribution of \ha\ EW as a function of clustercentric radius for all
galaxies as well as just the late--type galaxies (with $C>0.4$).  We have
performed a KS test between the late--type galaxies ($C>0.4$) in the tails
($>5$\AA) of the four \ha\ EW distributions in Figure \ref{histo} (the black
filled histograms) and the late--type galaxies in the field ({\it i.e.}, also
with $C>0.4$ and $>5$\AA). This analysis indicates that the inner three
clustercentric radii ($0.3R_{v}$, $0.9R_{v}$, $1.6R_{v}$) are inconsistent
with the field population, at the 82.5\%, 71\% and 99.8\% probability levels,
respectively. This suggests that the distribution of \ha\ EWs for strongly
star--forming ($>5$\AA) late--type galaxies does change with density, which is
consistent with our results in Section \ref{densityanalysis} and with those of
\citet{Hash98}.

\subsection{Physical Interpretation}\label{sec-physinterp}
The most striking result of this paper is the critical density (or radius)
where the SFR of galaxies changes from that of the field. This happens at a
density of $\sim 1\,h^{-2}_{75}\,{\rm Mpc^{-2}}$, which corresponds to a
clustercentric distance of between $2$ and $3$ virial radii.  In this section,
we attempt to understand this observation within the context of hierarchical
structure formation.  In hierarchical models, clusters and groups form from
slightly overdense regions in the initial matter distribution.  At first,
clusters expand with the Hubble flow until they reach a maximum size and then
de--couple from the expansion and collapse to form a virialized object.  Using
the spherical collapse model \citep{GG}, one can show that the point of
maximal size, also called the ``turn--around'' radius ($R_t$), is equal to a
density of approximately six times the mean density of the background at that
time.  The virial radius ($R_{v}$) corresponds to a radius where the density
is a few hundred times the mean density of the background. Therefore, there
are two natural scales for an overdensity in the Universe -- the virial radius
and the turn--around radius -- and the latter radius marks the limit of the
gravitational influence of a cluster or group. The region between the two
scales is known as the infall region.

We can estimate the ratio of virial ($R_{v}$) and turn--around ($R_t$) radius
as follows.  Suppose that the region between $R_{v}$ and $R_{t}$ is
completely empty.  Then $M_t/R_t^3\,=\, M_{v}/R_{v}^3\,(R_{v}/R_t)^3$, where
$M_t$ and $M_{v}$ are the masses of the cluster within the turn--around and
virial radii, respectively. Using $M_t/R_t^3 \sim 6$ and $M_{v}/R_{v}^3 \sim
300$, we find $R_t\simeq4\,R_{v}$.  In practice however, the region between
$R_{v}$ and $R_t$ is not empty and a more realistic case is to assume
that clusters or groups of galaxies follow, for example, the NFW mass
density profile \citep{NFW3}. We can thus find the turn--around radius by
integrating the profile until $M(<R)/R^3 = 6$. In this case, we find that
$R_t\simeq5\,R_{v}$, which is in good agreement with our naive calculation
above.

It is intriguing that this scale ($R_t\simeq5\,R_{v}$) is comparable to the
scale at which the cluster SFR becomes statistically indistinguishable from
the field.  However, it is unclear that this radius has any physical meaning
that might effect the SFR of individual galaxies, as the mechanisms proposed
for changing the SFR of galaxies in the cores of rich clusters, {\it e.g.,}
ram-pressure stripping of the gas, galaxy harassment \citep{harass}, tidal
disruption \citep{BV}, are unlikely to be important at low densities and/or
large virial radii. A more appropriate mechanism for changing the SFR of these
galaxies may be the merger or close tidal interactions of galaxies in less
dense groups within the infall regions of clusters
\citep{Z+96,ZM98,Kodama_cl0939}.

To obtain a more detailed physical understanding of our results, we must
compare our observations with numerical simulations. For example,
\citet{infall} and \citet{Diaferio} have recently showed that simple
(heuristic) models for the gas depletion of galaxies within hierarchical
models of structure formation are able to accurately reproduce the decrease
in the star--formation rate of galaxies seen within the cores of CNOC1
clusters of galaxies \citep{B+97,B+98}.  In both these works, the SFR of a
galaxy is affected by both local and external processes.  Locally, the SFR is
governed solely by the consumption rate of cold gas in the disk, dependent
only on gas density and the feedback model.  The only external process that
effects the SFR of a single galaxy is the stripping of its hot gas reservoir
after it merges with a larger ({\it e.g.,} group or cluster) halo.  Following
this, the SFR declines gradually, as the galaxy consumes the remaining cold,
disk gas.  Therefore, the main physical properties of a galaxy in the
simulation that controls these processes are the amount of cool gas in the
galaxy and the time since the last interaction with a larger halo.
\citet{Diaferio} predicts that the mean SFR of galaxies in clusters should be
lower than the field out to $2R_{v}$ (here $R_{v}$ is a 3--dimensional radius from
the simulations, while throughout the paper, we have quoted $R_{v}$ as a
projected radius from the cluster cores). This reduction is due to galaxies
that had been near the cluster core, but thrown out to large radii during
major mergers \citep{infall,EG02}. Our results are qualitatively similar to
the predictions of \citet{infall} and \citet{Diaferio} in that these
hierarchical models of structure formation can affect the SFR of galaxies
beyond the virial radius. However, we will require further comparison of the
simulations and observations to determine if these simple models are all that
is required to explain the data, or whether we need additional physical
processes that effect the SFR of galaxies at larger radii in the infall
regions of clusters, {\it i.e.,} from 2$R_{v}$ to 4$R_{v}$.

\subsection{Comparison with Previous Work}
\label{2df}
A similar analysis has been carried out using the LCRS
\citep{Hash98} and 2dF Galaxy Redshift Survey (2dFGRS) data \citep{2dF-sfr}.
For the 2dFGRS, the luminosity limits, redshift limits, and \ha\ EW
distributions are comparable with our sample, and both studies are
remarkably consistent in their main conclusions.  First, star formation is
reduced, relative to the field, out to $\gtrsim 3$ virial radii from known
clusters and groups.  Secondly, there appears to be a critical density of $1$
galaxy (brighter than $M^{\ast}+1.5$) per Mpc$^{2}$, below which there is a
weaker correlation with star-formation rate.

Our work moves beyond the 2dF study in several ways.  Most importantly, we
have computed a local density for every galaxy in the EDR, regardless of its
proximity to a cluster.  Moreover, our cluster catalog is dominated by systems
with much lower velocity dispersions than those in \citet{2dF-sfr}, so it
further emphasizes that the density--SFR relation is universal and does not
depend on the large-scale mass of the embedding structure. Furthermore,
\citet{2dF-sfr} are unable to compute H$\alpha$ fluxes, and hence star
formation rates, due to uncertainty in the continuum fluxing. This is not a
problem for the SDSS, and we have been able to show that there exists an
absolute density--SFR relation, as well as a relative one.  We have also used
the fluxes from another line (\oii) to demonstrate that our results are
robust to the effects of reddening.  Finally, the data analysed by
\citet{2dF-sfr} are preliminary in the sense that sampling is not complete
over their full extracted regions.  The consistency between our results shows
that this is not a large problem.

Recently, \citet{Kodama_cl0939} reported the detection of a ``break'' in the
colors of galaxies around distant cluster Abell 851 ($z=0.41$). They observed
an abrupt change in the colors of galaxies at a local surface density of
galaxies of $\sim 200\,{\rm h_{75}^{-2}\,Mpc^{-2}}$, which corresponds to a
radial distance of $\sim 1\,{\rm h_{75}^{-1}\,Mpc}$ from the center of this
cluster. Even after correcting for the differences in the luminosity limits of
the two data sets ($0.025L^{\star}_{V}$ in \citet{Kodama_cl0939} compared to
$0.4L^{\star}_{r^*}$ here), the ``break'' seen by \citet{Kodama_cl0939} occurs
at an order of magnitude greater surface density, and also a smaller
clustercentric radius, than the ``break'' reported here. Further observations
will be needed to determine if this difference is physical ({\it i.e.}, due to
differences in the redshift or luminosity limits of the two studies) or an
artifact of the different analyses.

\subsection{Future Work}

As the SDSS database increases, we will be able to increase the size and depth
of the sample used herein, {\it i.e.,} to increase the number of galaxy
clusters, as well as to relax some of the conservative selection criteria we
have imposed on our present sample. Furthermore, we will use additional
information on these galaxies extracted from both the SDSS photometry and
spectra.  For example, we can use the five passband SDSS photometry to look
for color gradients within these galaxies to fully correct for the aperture
biases as well as study how the environment effects the star formation {\it
within} a galaxy \citep[see][]{MW,MW00,Rose01}. We must also extend our study
to characterize the effects of the environment on the overall star formation
history of galaxies. For instance, we should include post--starburst galaxies
\citep[E+A, k+a and a+k; see ][]{Z+96,P+99,PSG,SDSS_Coma}, because these may
be key to fully understanding star--formation activity within dense
environments. In addition, we should probe fainter luminosities to determine
the rate of change of SFR per unit stellar luminosity. 

Most importantly, the research discussed in Section \ref{DMchap} clearly
requires a more robust, automated morphology for each of the SDSS galaxies
similar to the bulge--to--disk decompositions used by others
\citep{B+98,lowlx-morph}. We can then perform a multi--variate analysis on the
galaxy population that treats the SFR, local density, morphology, mass and
luminosity of the galaxies in a self--consistent manner. We must also expand
our definition of local density to include more information on the
higher--order moments of the galaxy distribution, {\it e.g.,} filaments.  A
full characterization of the effect of the environment on the overall star
formation history of galaxies is paramount to our understanding of the
physical processes that could be responsible for galaxy
evolution. 

\newpage

\section{Conclusions}\label{conclusions}
We have investigated the effects of the local galaxy environment on the
star--formation of galaxies using the Early Data Release of the Sloan Digital
Sky Survey. For this work, we have restricted our analysis to a
volume--limited sample of 8598 galaxies brighter than $M(r^*) \leq -20.45$
(k--corrected, for ${\rm H_0} = 75\,{\rm km\, s^{-1}\, Mpc^{-1}}$) over the
redshift range $0.05\le~z\le 0.095$. For all galaxies, we have characterized
the star--formation using the \ha\ and \oii\ EWs, as well as by computing their
star--formation rates (SFR) and normalized star--formation rates (SFRN; SFR
per unit luminosity) from the SDSS data. We have quantified the local galaxy
environment using the projected distance to the 10$^{th}$ nearest neighbor
brighter than $M^{\star}+1$ as well as by measuring the clustercentric
distance of galaxies from the cores of known clusters and groups of galaxies
(see Appendix).  We have extensively tested our results as discussed in detail
in the Appendix.

This work expands upon previous studies of the SFR of galaxies as a function of
environment in four ways: {\it i)} We extend such studies into the group and
poor cluster regime; {\it ii)} We focus on the low redshift universe; {\it
iii)} We have traced the SFR activity of galaxies well beyond the cores of the
clusters and groups into the field population; and {\it iv)} We are able to
accurately quantify the local galaxy density in a uniform manner that is not
subject to statistical background correction. We conclude that:

\begin{itemize}

\item The distributions of \oii\ and \ha\ EWs, as quantified by the median and
25$^{th}$ and 75$^{th}$ percentiles, changes as a function of local galaxy
density as measured using the distance to the $10^{th}$ nearest neighbor.  We
witness similar relations in the SFR and SFRN of galaxies as shown in Figure
\ref{densitysfr}. This effect is characterized in three ways.  First, there is
shift in the overall distributions of EW, SFR, and SFRN to lower values with
increasing galaxy density. Secondly, the skewness of the distributions
decreases with increasing local galaxy density, {\it i.e.,} the tail of
strongly star--forming galaxies (\ha\ EW $>5$\AA), as quantified by the
75$^{th}$ percentile of the distribution, is noticeably decreased in high
density regions. Finally, we see a ``break'' (or characteristic scale) in the
correlation between SFR (and SFRN) and density at a local galaxy density of
$\simeq 1\,h_{75}^{-2}\, {\rm Mpc^{-2}}$ (for galaxies brighter than
$M^\ast$).  Figures \ref{densityew} and \ref{densitysfr} represent the
density--SFR relation of galaxies that is analogous to the density--morphology
relation of galaxies (Dressler 1980).

\item The distribution of EW, SFR, and SFRN of galaxies, as quantified by the
median and 25$^{th}$ and 75$^{th}$ percentiles, changes as a function of
clustercentric radius for the $17$ clusters and groups discussed in this paper
(see Table \ref{tabclus}). This effect is most noticeable for the strongly
star--forming galaxies in the 75$^{th}$ percentiles of the distribution. Using
a KS test and boot--strapping re-sampling techniques, we find that the star
formation rates of galaxies begin to decrease, compared with field galaxies,
starting at 3-4 virial radii (with $>1\sigma$ statistical significance).
Within one virial radius of our clusters, the means (medians) of the \ha\ and
\oii\ EW distributions are 3.4\AA\ ($-0.7$\AA) and 3.2\AA\ (1.8\AA),
respectively.

\item As shown in Figure \ref{map1}, the break or characteristic density seen
at a (projected) galaxy density of $\simeq 1\,h_{75}^{-2}\,$ Mpc in the
density--SFR relation corresponds to $\sim 2$ and $3$ virial radii for the
systems discussed in this paper. Therefore, there is good agreement between
our results from the clusters and groups (discussed in Section 3.2) and the
general density--SFR relation (discussed in Section 3.1).

\item We have investigated the possible degeneracy between our density--SFR
relation and the density--morphology relation (Dressler 1980). There is some
evidence that the SFR of galaxies in dense regions varies even among galaxies
of the same morphology. For example, we have used the (inverse) concentration
index of SDSS galaxies to show that the tail of the strongly star--forming
(\ha\ EW $>5$\AA) galaxies observed in the \ha\, EW, SFR, and SFRN
distributions is dominated ($>75$\%) by late--type (spiral) galaxies. Using a
KS test, we have shown that this tail of late--type galaxies is different in
dense regions (within 2 virial radii) compared with similar galaxies in the
field. We stress, however, that the significance of these results remains
uncertain at present due to potential systematic biases in the morphological
classifications used herein.

\item Our results are qualitatively consistent with the predictions of
\citet{infall} and \citet{Diaferio} in that these hierarchical models of
structure formation can affect the SFR of galaxies well beyond the virial
radius. However, we require more detailed comparisons between the simulations
and observations to determine if such simple (heuristic) models are all that
is necessary to explain the data, or whether we need additional processes that
affect the SFR of galaxies at large clustercentric radii and/or low densities.

\item 
Our results are in good agreement with the recent 2dF work of Lewis et
al. (2002) as well as consistent with previous observations of a decrease in
the SFR of galaxies in the cores of distant clusters (Balogh et al. 1997;
1998). Taken together, these results demonstrate that the decrease in SFR of
galaxies in dense environments is a universal phenomenon over a wide range in
densities (from the rarefied field to poor groups to rich clusters) and
redshifts.

\end{itemize}



\begin{thebibliography}{}

\bibitem[Abell, Corwin, \& Olowin (1989)]{abell} Abell, G.~O., 
Corwin, H.~G., \& Olowin, R.~P.\ 1989, \apjs, 70, 1. 

\bibitem[{{B{\" o}hringer} {et~al.}(2001){B{\" o}hringer}, {Schuecker},
  {Guzzo}, {Collins}, {Voges}, {Schindler}, {Neumann}, {Cruddace}, {De Grandi},
  {Chincarini}, {Edge}, {MacGillivray}, \& {Shaver}}]{B+01}
{B{\" o}hringer}, H., {Schuecker}, P., {Guzzo}, L., {Collins}, C.~A., {Voges},
  W., {Schindler}, S., {Neumann}, D.~M., {Cruddace}, R.~G., {De Grandi}, S.,
  {Chincarini}, G., {Edge}, A.~C., {MacGillivray}, H.~T., \& {Shaver}, P. 2001,
  A\&A, 369, 826


\bibitem[{Balogh {et~al.}(1997)Balogh, Morris, Yee, Carlberg, \&
  Ellingson}]{B+97}
Balogh, M.~L., Morris, S.~L., Yee, H. K.~C., Carlberg, R.~G., \& Ellingson, E.
  1997, ApJL, 488, 75

\bibitem[{Balogh {et~al.}(1999)Balogh, Morris, Yee, Carlberg, \&
  Ellingson}]{PSG}
---. 1999, ApJ, 527, 54


\bibitem[{Balogh {et~al.}(1998)Balogh, Schade, Morris, Yee, Carlberg, \&
  Ellingson}]{B+98}
Balogh, M.~L., Schade, D., Morris, S.~L., Yee, H. K.~C., Carlberg, R.~G., \&
  Ellingson, E. 1998, ApJL, 504, 75

\bibitem[{{Balogh} {et~al.}(2000){Balogh}, {Navarro}, \& {Morris}}]{infall}
{Balogh}, M.~L., {Navarro}, J.~F., \& {Morris}, S.~L. 2000, ApJ, 540, 113

\bibitem[{{Balogh} {et~al.}(2002{\natexlab{a}}){Balogh}, {Couch}, {Smail},
  {Bower}, \& {Glazebrook}}]{A1689}
{Balogh}, M.~L., {Couch}, W.~J., {Smail}, I., {Bower}, R.~G., \& {Glazebrook}.
  2002{\natexlab{a}}, MNRAS, submitted

\bibitem[{{Balogh} {et~al.}(2002{\natexlab{b}}){Balogh}, {Smail}, {Bower},
  {Ziegler}, {Smith}, {Davies}, {Gaztelu}, {Kneib}, \& {Ebeling}}]{lowlx-morph}
{Balogh}, M.~L., {Smail}, I., {Bower}, R.~G., {Ziegler}, B.~L., {Smith}, G.~P.,
  {Davies}, R.~L., {Gaztelu}, A., {Kneib}, J.-P., \& {Ebeling}, H.
  2002{\natexlab{b}}, ApJ, 566, 123

\bibitem[{{Bartholomew} {et~al.}(2001){Bartholomew}, {Rose}, {Gaba}, \&
  {Caldwell}}]{Bart01}
{Bartholomew}, L.~J., {Rose}, J.~A., {Gaba}, A.~E., \& {Caldwell}, N. 2001, AJ,
  122, 2913

\bibitem[{Baugh {et~al.}(1996)Baugh, Cole, \& Frenk}]{semianal}
Baugh, C.~M., Cole, S., \& Frenk, C.~S. 1996, MNRAS, 283, 1361

\bibitem[{Bekki {et~al.}(2002)}]{bekki}
Bekki, K., Shioya, Y.,  Couch, W. J., 2002, ApJ, accepted (astro-ph/0206207)

\bibitem[{{Bernstein} {et~al.}(1995){Bernstein}, {Nichol}, {Tyson}, {Ulmer}, \&
  {Wittman}}]{B+95}
{Bernstein}, G.~M., {Nichol}, R.~C., {Tyson}, J.~A., {Ulmer}, M.~P., \&
  {Wittman}, D. 1995, AJ, 110, 1507

\bibitem[{{Blanton} {et~al.}(2001){Blanton}, {Dalcanton}, {Eisenstein},
  {Loveday}, {Strauss}, {SubbaRao}, {Weinberg}, \& {the Sloan
  collaboration}}]{Sloan_lf}
{Blanton}, M.~R., {Dalcanton}, J., {Eisenstein}, D., {Loveday}, J., {Strauss},
  M.~A., {SubbaRao}, M., {Weinberg}, D.~H., \& {the Sloan collaboration}. 2001,
  AJ, 121, 2358

\bibitem[{Byrd \& Valtonen(1990)}]{BV}
Byrd, G. \& Valtonen, M. 1990, ApJ, 350, 89

\bibitem[{{Castander} {et~al.}(2001){Castander}, {Nichol}, {Merrelli},
  {Burles}, {Pope}, {Connolly}, {Uomoto}, {Gunn}, {Anderson}, {Annis},
  {Bahcall}, {Boroski}, {Brinkmann}, {Carey}, {Crocker}, {Csabai}, {Doi},
  {Frieman}, {Fukugita}, {Friedman}, {Hilton}, {Hindsley}, {Ivezi{\' c}},
  {Kent}, {Lamb}, {Leger}, {Long}, {Loveday}, {Lupton}, {MacGillivray},
  {Meiksin}, {Munn}, {Newcomb}, {Okamura}, {Owen}, {Pier}, {Rockosi},
  {Schlegel}, {Schneider}, {Seigmund}, {Smee}, {Snir}, {Starkman}, {Stoughton},
  {Szokoly}, {Stubbs}, {SubbaRao}, {Szalay}, {Thakar}, {Tremonti}, {Waddell},
  {Yanny}, \& {York}}]{SDSS_Coma}
{Castander}, F.~J., {Nichol}, R.~C., {Merrelli}, A., {Burles}, S., {Pope}, A.,
  {Connolly}, A.~J., {Uomoto}, A., {Gunn}, J.~E., {Anderson}, J.~E., {Annis},
  J., {Bahcall}, N.~A., {Boroski}, W.~N., {Brinkmann}, J., {Carey}, L.,
  {Crocker}, J.~H., {Csabai}, I.~., {Doi}, M., {Frieman}, J.~A., {Fukugita},
  M., {Friedman}, S.~D., {Hilton}, E.~J., {Hindsley}, R.~B., {Ivezi{\' c}}, {\v
  Z}., {Kent}, S., {Lamb}, D.~Q., {Leger}, R.~F., {Long}, D.~C., {Loveday}, J.,
  {Lupton}, R.~H., {MacGillivray}, H., {Meiksin}, A., {Munn}, J.~A., {Newcomb},
  M., {Okamura}, S., {Owen}, R., {Pier}, J.~R., {Rockosi}, C.~M., {Schlegel},
  D.~J., {Schneider}, D.~P., {Seigmund}, W., {Smee}, S., {Snir}, Y.,
  {Starkman}, L., {Stoughton}, C., {Szokoly}, G.~P., {Stubbs}, C., {SubbaRao},
  M., {Szalay}, A., {Thakar}, A.~R., {Tremonti}, C., {Waddell}, P., {Yanny},
  B., \& {York}, D.~G. 2001, AJ, 121, 2331

\bibitem[{{Charlot} {et~al.}(2001){Charlot}, {Kauffmann}, {Longhetti},
  {Tresse}, {White}, {Maddox}, \& {Fall}}]{CL_ext}
{Charlot}, S.~., {Kauffmann}, G., {Longhetti}, M., {Tresse}, L., {White}, S.
  D.~M., {Maddox}, S.~J., \& {Fall}, S.~M. 2001, MNRAS, in press,
  astroph/111289

\bibitem[{{Charlot} \& {Longhetti}(2001)}]{CL}
{Charlot}, S.~. \& {Longhetti}, M. 2001, MNRAS, 323, 887

\bibitem[{{Cole} {et~al.}(2000){Cole}, {Lacey}, {Baugh}, \& {Frenk}}]{Cole2000}
{Cole}, S., {Lacey}, C.~G., {Baugh}, C.~M., \& {Frenk}, C.~S. 2000, MNRAS, 319,
  168


\bibitem[{{Couch} {et~al.}(2001){Couch}, {Balogh}, {Bower}, {Smail},
  {Glazebrook}, \& {Taylor}}]{C+01}
{Couch}, W.~J., {Balogh}, M.~L., {Bower}, R.~G., {Smail}, I., {Glazebrook}, K.,
  \& {Taylor}, M. 2001, ApJ, 549, 820

\bibitem[{{Diaferio} {et~al.}(2001){Diaferio}, {Kauffmann}, {Balogh}, {White},
  {Schade}, \& {Ellingson}}]{Diaferio}
{Diaferio}, A., {Kauffmann}, G., {Balogh}, M.~L., {White}, S. D.~M., {Schade},
  D., \& {Ellingson}, E. 2001, MNRAS, 323, 999

\bibitem[{Dressler(1980)}]{Dressler}
Dressler, A. 1980, ApJ, 236, 351

\bibitem[{Dressler {et~al.}(1997)Dressler, Oemler, Couch, Smail, Ellis, Barger,
  Butcher, Poggianti, \& Sharples}]{D+97}
Dressler, A., Oemler, A., Couch, W.~J., Smail, I., Ellis, R.~S., Barger, A.,
  Butcher, H.~R., Poggianti, B.~M., \& Sharples, R.~M. 1997, ApJ, 490, 577

\bibitem[{{Evrard} \& {Gioia}(2002)}]{EG02}
{Evrard}, A.~E. \& {Gioia}, I. 2002, ApJ, in preparation


\bibitem[{{Fukugita} {et~al.}(1996){Fukugita}, {Ichikawa}, {Gunn}, {Doi},
  {Shimasaku}, \& {Schneider}}]{F+96}
{Fukugita}, M., {Ichikawa}, T., {Gunn}, J.~E., {Doi}, M., {Shimasaku}, K., \&
  {Schneider}, D.~P. 1996, AJ, 111, 1748

\bibitem[{{Girardi} {et~al.}(1998){Girardi}, {Giuricin}, {Mardirossian},
  {Mezzetti}, \& {Boschin}}]{Girardi98}
{Girardi}, M., {Giuricin}, G., {Mardirossian}, F., {Mezzetti}, M., \&
  {Boschin}, W. 1998, ApJ, 505, 74

\bibitem[{{Gladders} \& {Yee}(2000)}]{GY00}
{Gladders}, M.~D. \& {Yee}, H.~K.~C. 2000, AJ, 120, 2148

\bibitem[{{Goto} {et~al.}(2002){Goto}, {Sekiguchi}, {Nichol}, Bahcall, Kim,
  Annis, Ivezic, Brinkmann, Hennessy, Szokoly, \& Tucker}]{tomo}
{Goto}, T., {Sekiguchi}, M., {Nichol}, R.~C., Bahcall, N.~A., Kim, R. S.~J.,
  Annis, J., Ivezic, Z., Brinkmann, J., Hennessy, G.~S., Szokoly, G.~P., \&
  Tucker, D.~L. 2002, AJ, in press, astroph/0112482

\bibitem[{{Gunn} \& {Gott}(1972)}]{GG}
{Gunn}, J.~E. \& {Gott}, J. R.~I. 1972, ApJ, 176, 1

\bibitem[Gunn et al. (1998)]{gunn} Gunn, J. E. et al. 1998, AJ, 116, 3040

\bibitem[{Hashimoto} {et~al.}(1998)]{Hash98} 
Hashimoto, Y., Oemler, A.~J., Lin, H., \& Tucker, D.~L.\ 1998, \apj, 499, 
589 

\bibitem[{{Heavens}(1993)}]{H93}
{Heavens}, A.~F. 1993, MNRAS, 263, 735

\bibitem[{{Hopkins} {et~al.}(2001){Hopkins}, {Connolly}, {Haarsma}, \&
  {Cram}}]{Hopkins01}
{Hopkins}, A.~M., {Connolly}, A.~J., {Haarsma}, D.~B., \& {Cram}, L.~E. 2001,
  AJ, 122, 288

\bibitem[Hogg et al. (1998)]{hogg}Hogg, D.~W., Schlegel, D.~J., Finkbeiner, D.~P., \& Gunn, J.~E. 2001, AJ, 122, 2129

\bibitem[{{Jansen} {et~al.}(2000){Jansen}, {Fabricant}, {Franx}, \&
  {Caldwell}}]{Jansen}
{Jansen}, R.~A., {Fabricant}, D., {Franx}, M., \& {Caldwell}, N. 2000, ApJS,
  126, 331

\bibitem[{{Kauffmann} {et~al.}(1993){Kauffmann}, {White}, \&
  {Guiderdoni}}]{KWG}
{Kauffmann}, G., {White}, S. D.~M., \& {Guiderdoni}, B. 1993, MNRAS, 264, 201

\bibitem[{Kennicutt(1983)}]{K83}
Kennicutt, R.~C. 1983, ApJ, 272, 54


\bibitem[{Kennicutt(1992)}]{K92}
---. 1992, ApJ, 388, 310

\bibitem[{{Kennicutt}(1998{\natexlab{a}})}]{Kenn_review}
{Kennicutt}, R.~C., J. 1998{\natexlab{a}}, ARA\&A, 36, 189

\bibitem[{{Kennicutt}(1998{\natexlab{b}})}]{K98}
{Kennicutt}, R.~C. 1998{\natexlab{b}}, ApJ, 498, 541


\bibitem[{{Kewley} {et~al.}(2001){Kewley}, {Dopita}, {Sutherland}, {Heisler},
  \& {Trevena}}]{K+01}
{Kewley}, L.~J., {Dopita}, M.~A., {Sutherland}, R.~S., {Heisler}, C.~A., \&
  {Trevena}, J. 2001, ApJ, 556, 121

\bibitem[{Konchanek} {et al}(2000)]{kon92}Kochanek, C.S., Pahre, M.A., Falco, E.E., 2000, submitted, see astro-ph/0011458

\bibitem[{{Kodama} {et~al.}(2001){Kodama}, {Smail}, {Nakata}, {Okamura}, \&
  {Bower}}]{Kodama_cl0939}
{Kodama}, T., {Smail}, I., {Nakata}, F., {Okamura}, S., \& {Bower}, R.~G. 2001,
  ApJL, 562, L9


\bibitem[{{Lewis} {et~al.}(2002){Lewis}, {Balogh}, {De Propris}, {Couch},
  {Bower}, {Offer}, {Bland-HAwthorn}, {Baldry}, {Baugh}, {Bridges}, {Cannon},
  {Cole}, {Colless}, {Collins}, {Cross}, {Dalton}, P., {Efstathiou}, {Ellis},
  {Frenk}, {Glazebrook}, {Hawkins}, {Jackson}, {Lahav}, {Lumsden}, {Maddox},
  {Madgwick}, {Norberg}, {Peacock}, {Percival}, {Peterson}, {Sutherland}, \&
  {Taylor}}]{2dF-sfr}
{Lewis}, I., {Balogh}, M., {De Propris}, R., {Couch}, W., {Bower}, R., {Offer},
  A., {Bland-HAwthorn}, J., {Baldry}, I.~K., {Baugh}, C., {Bridges}, T.,
  {Cannon}, R., {Cole}, S., {Colless}, M., {Collins}, C., {Cross}, N.,
  {Dalton}, G., P., D.~S., {Efstathiou}, G., {Ellis}, R.~S., {Frenk}, C.~S.,
  {Glazebrook}, K., {Hawkins}, E., {Jackson}, C., {Lahav}, O., {Lumsden}, S.,
  {Maddox}, S., {Madgwick}, D., {Norberg}, P., {Peacock}, J.~A., {Percival},
  W., {Peterson}, B.~A., {Sutherland}, W., \& {Taylor}, K. 2002, MNRAS, in
  press

\bibitem[{{Lilly} {et~al.}(1996){Lilly}, {Le Fevre}, {Hammer}, \&
  {Crampton}}]{L96}
{Lilly}, S.~J., {Le Fevre}, O., {Hammer}, F., \& {Crampton}, D. 1996, ApJL,
  460, L1

\bibitem[{{Lupton} {et~al.}(2001){Lupton}, {Gunn}, {Ivezi{\' c}}, {Knapp},
  {Kent}, \& {Yasuda}}]{SDSS_software}
{Lupton}, R.~H., {Gunn}, J.~E., {Ivezi{\' c}}, Z., {Knapp}, G.~R., {Kent}, S.,
  \& {Yasuda}, N. 2001, in Astronomical Data Analysis Software and Systems X,
  ASP Conference Proceedings, Vol. 238. Edited by F. R. Harnden, Jr., Francis
  A. Primini, and Harry E. Payne. San Francisco: Astronomical Society of the
  Pacific, ISSN: 1080-7926, 2001., p.269, Vol.~10, 269

\bibitem[{{Lupton} {et~al.}(1999){Lupton}, {Gunn}, \& {Szalay}}]{LGS}
{Lupton}, R.~H., {Gunn}, J.~E., \& {Szalay}, A.~S. 1999, AJ, 118, 1406


\bibitem[{{Magliocchetti} {et~al.}(2000){Magliocchetti}, {Maddox}, {Wall},
  {Benn}, \& {Cotter}}]{FIRST}
{Magliocchetti}, M., {Maddox}, S.~J., {Wall}, J.~V., {Benn}, C.~R., \&
  {Cotter}, G. 2000, MNRAS, 318, 1047

\bibitem[{{Mahdavi} \& {Geller}(2001)}]{MG01}
{Mahdavi}, A. \& {Geller}, M.~J. 2001, ApJL, 554, L129

\bibitem[{{Miller} {et~al.}(2001){Miller}, {Genovese}, {Nichol}, {Wasserman},
  {Connolly}, {Reichart}, {Hopkins}, {Schneider}, \& {Moore}}]{Miller01}
{Miller}, C.~J., {Genovese}, C., {Nichol}, R.~C., {Wasserman}, L., {Connolly},
  A., {Reichart}, D., {Hopkins}, A., {Schneider}, J., \& {Moore}, A. 2001, AJ,
  122, 3492

\bibitem[{{Moore} {et~al.}(1999){Moore}, {Lake}, {Quinn}, \& {Stadel}}]{harass}
{Moore}, B., {Lake}, G., {Quinn}, T., \& {Stadel}, J. 1999, MNRAS, 304, 465

\bibitem[{{Moss} \& {Whittle}(1993)}]{MW}
{Moss}, C. \& {Whittle}, M. 1993, ApJL, 407, L17

\bibitem[{{Moss} \& {Whittle}(2000)}]{MW00}
---. 2000, MNRAS, 317, 667

\bibitem[{Navarro {et~al.}(1996)Navarro, Frenk, \& White}]{NFW3}
Navarro, J.~F., Frenk, C.~S., \& White, S. D.~M. 1996, ApJ, 462, 563

\bibitem[Nichol et al.(2001)]{2001misk.conf..613N} Nichol, R.~C.~et al.\ 
2001, Mining the Sky, 613. 



\bibitem[Pier et al. (2002)]{peir} Pier, J. R., Munn, J. A., Hindsley, R. B., Hennessy, G. S., Kent, S. M., Lupton, R. H., \& Ivezic, Z. 2002, AJ, submitted 

\bibitem[{{Pimbblet} {et~al.}(2001){Pimbblet}, {Smail}, {Kodama}, {Couch},
  {Edge}, {Zabludoff}, \& {O'Hely}}]{kap2}
{Pimbblet}, K.~A., {Smail}, I., {Kodama}, T., {Couch}, W.~J., {Edge}, A.~C.,
  {Zabludoff}, A.~I., \& {O'Hely}, E. 2001, MNRAS, in press

\bibitem[{Poggianti {et~al.}(1999)Poggianti, Smail, Dressler, Couch, Barger,
  Butcher, Ellis, \& Oemler}]{P+99}
Poggianti, B.~M., Smail, I., Dressler, A., Couch, W.~J., Barger, A.~J.,
  Butcher, H., Ellis, R.~S., \& Oemler, A. 1999, ApJ, 518, 576

\bibitem[{Postman \& Geller(1984)}]{PG84}
Postman, M. \& Geller, M.~J. 1984, ApJ, 281, 95

\bibitem[{{Postman} {et~al.}(2001){Postman}, {Lubin}, \& {Oke}}]{PLO}
{Postman}, M., {Lubin}, L.~M., \& {Oke}, J.~B. 2001, AJ, 122, 1125

\bibitem[{{Quilis} {et~al.}(2000){Quilis}, {Moore}, \& {Bower}}]{QMB}
{Quilis}, V., {Moore}, B., \& {Bower}, R. 2000, Science, 288, 1617

\bibitem[{{Rose} {et~al.}(2001){Rose}, {Gaba}, {Caldwell}, \&
  {Chaboyer}}]{Rose01}
{Rose}, J.~A., {Gaba}, A.~E., {Caldwell}, N., \& {Chaboyer}, B. 2001, AJ, 121,
  793

\bibitem[{Salpeter(1955)}]{Sp}
Salpeter, E.~E. 1955, ApJ, 121, 161

\bibitem[{{Schlegel} {et~al.}(1998){Schlegel}, {Finkbeiner}, \& {Davis}}]{SFD}
{Schlegel}, D.~J., {Finkbeiner}, D.~P., \& {Davis}, M. 1998, ApJ, 500, 525

\bibitem[{{Shane} \& {James}(2001)}]{SJ}
{Shane}, N.~S. \& {James}, P.~A. 2001, astro-ph/, 0112183

\bibitem[{{Shimasaku} {et~al.}(2001){Shimasaku}, {Fukugita}, {Doi}, {Hamabe},
  {Ichikawa}, {Okamura}, {Sekiguchi}, {Yasuda}, {Brinkmann}, {Csabai},
  {Ichikawa}, {Ivezi{\' c}}, {Kunszt}, {Schneider}, {Szokoly}, {Watanabe}, \&
  {York}}]{SDSS_morph}
{Shimasaku}, K., {Fukugita}, M., {Doi}, M., {Hamabe}, M., {Ichikawa}, T.,
  {Okamura}, S., {Sekiguchi}, M., {Yasuda}, N., {Brinkmann}, J., {Csabai},
  I.~., {Ichikawa}, S., {Ivezi{\' c}}, Z., {Kunszt}, P.~Z., {Schneider}, D.~P.,
  {Szokoly}, G.~P., {Watanabe}, M., \& {York}, D.~G. 2001, AJ, 122, 1238

\bibitem[Smith et al. (2002)]{smith02} Smith, J.~A. et al. 2002, AJ, 123, 2121

\bibitem[{{Somerville} \& {Primack}(1999)}]{SP99}
{Somerville}, R.~S. \& {Primack}, J.~R. 1999, MNRAS, 310, 1087


\bibitem[{{Stickel} {et~al.}(2002){Stickel}, {Klaas}, {Lemke}, \&
  {Mattila}}]{SKLD}
{Stickel}, M., {Klaas}, K., {Lemke}, D., \& {Mattila}, K. \ 2002, \aap, 383, 367 

\bibitem[Stoughton et al.(2002)]{stoughton} Stoughton, C.~et al.\ 
2002, \aj, 123, 485. 

\bibitem[Strauss et al.(2002)]{strauss02} Strauss, M.~et al.\
2002, \aj, submitted

\bibitem[{{Sullivan} {et~al.}(2001){Sullivan}, {Mobasher}, {Chan}, {Cram},
  {Ellis}, {Treyer}, \& {Hopkins}}]{S+01}
{Sullivan}, M., {Mobasher}, B., {Chan}, B., {Cram}, L., {Ellis}, R., {Treyer},
  M., \& {Hopkins}, A. 2001, ApJ, 558, 72

\bibitem[{{Voges} {et~al.}(2000){Voges}, {Aschenbach}, {Boller}, {Brauninger},
  {Briel}, {Burkert}, {Dennerl}, {Englhauser}, {Gruber}, {Haberl}, {Hartner},
  {Hasinger}, {Pfeffermann}, {Pietsch}, {Predehl}, {Schmitt}, {Trumper}, \&
  {Zimmermann}}]{voges}
{Voges}, W., {Aschenbach}, B., {Boller}, T., {Brauninger}, H., {Briel}, U.,
  {Burkert}, W., {Dennerl}, K., {Englhauser}, J., {Gruber}, R., {Haberl}, F.,
  {Hartner}, G., {Hasinger}, G., {Pfeffermann}, E., {Pietsch}, W., {Predehl},
  P., {Schmitt}, J., {Trumper}, J., \& {Zimmermann}, U. 2000, VizieR Online
  Data Catalog, 9029, 0

\bibitem[{Whitmore {et~al.}(1993)Whitmore, Gilmore, \& Jones}]{WGJ}
Whitmore, B.~C., Gilmore, D.~M., \& Jones, C. 1993, ApJ, 407, 489

\bibitem[{{York} {et~al.}(2000){York}, {Adelman}, {Anderson}, {Anderson},
  {Annis}, {Bahcall}, {Bakken}, {Barkhouser}, \& {the Sloan
  collaboration}}]{SDSS_tech_short}
{York}, D.~G., {Adelman}, J., {Anderson}, J.~E., {Anderson}, S.~F., {Annis},
  J., {Bahcall}, N.~A., {Bakken}, J.~A., {Barkhouser}, R., \& {the Sloan
  collaboration}. 2000, AJ, 120, 1579


\bibitem[{Zabludoff {et~al.}(1996)Zabludoff, Zaritsky, Lin, Tucker, Hashimoto,
  Shectman, Oemler, \& Kirshner}]{Z+96}
Zabludoff, A.~I., Zaritsky, D., Lin, H., Tucker, D., Hashimoto, Y., Shectman,
  S.~A., Oemler, A., \& Kirshner, R.~P. 1996, ApJ, 466, 104

\bibitem[{{Zabludoff} \& {Mulchaey}(1998)}]{ZM98}
{Zabludoff}, A.~I. \& {Mulchaey}, J.~S. 1998, ApJ, 496, 39


\bibitem[Zaritsky, Zabludoff, \& Willick(1995)]{1995AJ....110.1602Z} 
Zaritsky, D., Zabludoff, A.~I., \& Willick, J.~A.\ 1995, \aj, 110, 1602. 



\end{thebibliography}
\section*{Acknowledgements}
We would like to thank Rupert Croft, Tim Heckman, David Schlegel, Larry
Wasserman, Christy Tremonti, Guinevere Kauffmann and Daniel Eisenstein for
useful discussions about this work. We thank the 2dFGRS team for sharing their
results with us prior to publication. We thank an anonymous referee for his
helpful comments which made this paper better.  PG and AKR acknowledge
financial support from the NASA LTSA program (grant number NAG5-7926). TG
acknowledges financial support from the Japan Society for the Promotion of
Science (JSPS) through JSPS Research Fellowships for Young Scientists.  MLB
acknowledges support from a PPARC rolling grant for extragalactic astronomy at
Durham. AIZ acknowledges support form the NASA LTSA program (grant number
NAG5-11108).  Funding for the creation and distribution of the SDSS Archive
has been provided by the Alfred P. Sloan Foundation, the Participating
Institutions, the National Aeronautics and Space Administration, the National
Science Foundation, the U.S. Department of Energy, the Japanese
Monbukagakusho, and the Max Planck Society. The SDSS Web site is
http://www.sdss.org/. The SDSS is managed by the Astrophysical Research
Consortium (ARC) for the Participating Institutions. The Participating
Institutions are The University of Chicago, Fermilab, the Institute for
Advanced Study, the Japan Participation Group, The Johns Hopkins University,
Los Alamos National Laboratory, the Max-Planck-Institute for Astronomy (MPIA),
the Max-Planck-Institute for Astrophysics (MPA), New Mexico State University,
Princeton University, the United States Naval Observatory, and the University
of Washington

\appendix


\subsection{Selection of the Groups and Clusters of Galaxies}\label{clus_sel}

A critical part of our study of galaxy star--formation activity in dense
environments is the selection of virialized groups and clusters of galaxies
within the EDR. This was achieved using the objective C4 algorithm described
in Nichol et al. (2001) and Miller et al. (2002, in prep.), which assumes that
galaxy groups and clusters have a co-evolving set of galaxies with similar
colors, {\it e.g.,} within the E/S0 ridgeline. Like any cluster--finding
algorithm, this does bias our selection of systems because we assume a model
for the objects that we are searching for. However, such biases can be
quantified using simulations (see Miller et al. 2002, in prep.) and is
preferred to using visually--compiled catalogs, like Abell \citep{abell},
whose selection function is hard to quantify through simulations.  We stress
that these criteria are used only to find the clusters and not to assign
membership for subsequent analysis.

We briefly outline the methodology we used herein which is based on the C4
algorithm discussed in more detail in Miller et al. (2002, in prep). For each
galaxy in our sample, we count the number of neighbors within a
seven-dimensional box (four colors, RA, DEC and a redshift). The size of the
box in the color dimensions is determined from the errors in the galaxy
magnitudes and the intrinsic scatter in galaxy colors within the cores of
groups and clusters. The angular size corresponds to 1\,h$_{100}^{-1}\,$Mpc at
the redshift of the galaxy. The size of the box in the redshift direction is
$\Delta z = 0.1$ and thus only rejects obvious foreground and background
galaxies.  We then compare this galaxy count (around each galaxy) to the
distribution of neighbor counts for the whole field population and calculate
the probability that the target galaxy is a field galaxy. The field
distribution is determined by moving the seven-dimensional box to 100 randomly
chosen galaxies with similar seeing and galactic extinction
measurements. Using the False Discovery Rate thresholding technique of
\citet{Miller01}, we then identify galaxies that have a low probability of
being in the field, {\it i.e.,} such that $< 5\%$ of these galaxies could be
mistaken field galaxies. We note that the C4 algorithm does not assume a given
color for these galaxies; it only assumes that these galaxies have similar
colors. This non-parametric approach significantly reduces the risk of biasing
our cluster sample because we are not modeling the expected colors of the
cluster galaxy population \citep{GY00,tomo}.  The details of the full C4
algorithm, including the selection function and completeness limits, are
presented in Miller et al. (2002, in prep.).

We first applied the above implementation of the C4 methodology to the EDR
spectroscopic data, {\it i.e.,} to all galaxies with a redshift. This finds
$\sim5\%$ of galaxies to be located in highly clustered regions (in our
seven-dimensional space), and it is therefore possible to locate candidate
clusters and groups within this highly--clustered galaxy dataset using a
non-parametric density estimator. For each candidate cluster, we then compute
the mean centroid and redshift of each system using all of the EDR data
available, {\it i.e.,} we now include all galaxies in the EDR spectroscopic
and photometric data sets regardless of the C4 algorithm. In addition, we also
estimate the line--of--sight velocity dispersion ($\sigma_r$) of each system
using a robust $\sigma_r$ estimator (see Miller et al. 2002 in prep.)  that
iteratively rejects all galaxies $> 3\sigma_r$ away from the mean redshift. We
then fit a Gaussian to the velocity distribution of each candidate cluster and
keep only systems for which the difference between the standard deviation
about the mean and the $1\sigma$ dispersion of the fitted Gaussian is less
than a factor of two. This approach removes spurious systems that account for
less than 10\% of whole sample and does not affect our results.

We present in Table \ref{tabclus} the $17$ groups and clusters of galaxies
that were found in the EDR via this implementation of the C4 algorithm and
that satisfy the selection criteria discussed above (in total, 30 clusters
were detected in the EDR, but $13$ of these were outside the redshift limits
used herein).  We provide the RA and DEC of the cluster centroids (in
degrees), their mean redshift, the line--of--sight velocity dispersion derived
from the Gaussian fit to the spectroscopic data ($\sigma_r(1)$), the
line--of--sight velocity dispersion derived from the standard deviation of the
galaxy distribution ($\sigma_r(2)$), the number of cluster members within 2
virial radii ($h = 0.75$) of the cluster center, brighter than $M_r = -20.45$,
and within $\pm 3\sigma_r(1)$ of the mean cluster redshift, and the X-ray
luminosity derived from the ROSAT All-Sky Survey \citep[RASS,][]{voges} data
in units of $10^{44}\, {\rm erg\,s^{-1}}$ (0.5-2.4 keV; $h=0.75$). The last
column of the table provides the name (and redshift if available) of the
nearest known cluster within 15 arcminutes from the NED database.  Details of
these systems are discussed in Miller et al. (2002, in prep.) along with
redshift histograms and color--magnitude relations that illustrate the reality
of these overdensities.

As can be seen from Table \ref{tabclus}, our C4 systems span a wide range of
velocity dispersions (from $\simeq 200\, {\rm km\,s^{-1}}$ to $ 1000\, {\rm
km\,s^{-1}}$) and X--ray luminosities (from $10^{42}\,h_{75}^{-2} {\rm
erg\,s^{-1}}$ to $\sim 4\times 10^{44}\,h_{75}^{-2} {\rm erg\,s^{-1}}$). The
relation between these two physical quantities is consistent with the ${\rm
L_x}$--$\sigma_r$ relation of \citet{ZM98} and \citet{MG01}, while the ratio
of groups to clusters in our sample is consistent with expectations based on
the volume of our galaxy sample, {\it i.e.,} in $\simeq 460\, {\rm deg^2}$ of
sky, we would only expect to find a few X--ray luminous clusters \citep{B+01}.
Therefore, our sample mostly covers the group and poor cluster regime
($\sim10^{43}\, h_{75}^{-2} {\rm erg\,s^{-1}}$ in X--ray luminosity). This is
different from most previous studies of environment-dependent star--formation
which have focused on rich clusters, and thus extends such studies to more
common galactic environments.

\subsection{SDSS Analysis Pipelines, Survey Strategy and Design}
\label{selectionbias}
As discussed in Stoughton et al. (2002) and Frieman et al. (in prep.), the
SDSS data analysis pipelines have been extensively tested and can accurately
determine the main characteristics of emission and absorption lines,
especially at the signal--to--noise ratios used herein.  
As part of a detailed check on these measurements, we have studied the 1231
duplicate observations of SDSS galaxies discussed in Section
\ref{selectioncriteria}. In Figure \ref{bernardi}, we show the percentage
difference between these duplicate observations of \ha\ EW as a function of
EW.  As expected, the measurement error is large for small EWs, but approaches
$\simeq10\%$ for \ha\ EW$\gtrsim 5$\AA. We further propagate these errors
through to our measurements of the SFR, and present in Table \ref{errors} the
mean and median percentage difference of the SFR as a function of SFR.  For
low SFRs, the error is substantial, but above a SFR of $\simeq1$\Msun$\, {\rm
yr^{-1}}$, the mean and median error is below 20\%.

\subsection{Stellar Absorption}
The present SPECTRO1D analysis pipeline does not account for possible stellar
absorption, and we have made no correction for this in our measurements of the
\ha\ emission line fluxes in this paper. This does not affect our results
because: {\it i)} The effect is expected to be $\lesssim 5$ \AA, which is
small relative to the strong emission-line galaxies in which we are
interested; {\it ii)} We include all galaxies (regardless of the size and sign
of their measured EW) when calculating the percentiles of the EW and SFR
distributions. If stellar absorption is only a weak function of EW or SFR,
then its effect will be to systematically shift the whole distribution to
smaller values that, in turn, will be reflected in the measured median,
$25^{th}$, and $75^{th}$ percentile of the EW distribution; {\it iii)} Our
results are consistent for both the \oii\ and \ha\ emission lines, which is
reassuring because the \oii\ emission lines are unaffected by stellar
absorption. Furthermore, \oii\ and \ha\ reside in different parts of the SDSS
spectrum (and were observed on different CCDs in the SDSS spectrographs) and
thus suffer different potential problems, {\it e.g.,} the determination of the
continuum near \oii\ can be severely affected by the $4000$\AA\
break. Therefore, the fact these two lines give similar answers suggests that
our results are not significantly affected by stellar absorption.

\subsection{Aperture Bias}
 Aperture bias is a concern because
we have observed all our galaxies through a fixed angular ($3$
arcseconds) fiber that is smaller than the angular size of our
galaxies (see Konchanek, Pahre \& Falco 2000 for a discussion of
possible effects of aperture bias).
Such a bias
could result in a systematic increase in the observed \ha\ EW and SFR
for higher redshift galaxies relative to lower redshift galaxies,
because more of the galaxy light is passing down the fiber. To
minimize this potential bias, we have followed the conclusions of
Zaritsky, Zabludoff \& Willick (1995) and restricted our sample of
galaxies to $z>0.05$. Above this redshift, Zaritsky et al. (1995)
showed that for the LCRS, the spectral classifications of LCRS
galaxies, which were observed through a $3.5$ arcsecond fiber, are
statistically unaffected by aperture bias.  In fact, the results of
Zaritsky et al. (1995) suggest that only $\sim10\%$ of spiral galaxies
in our sample are mis--classified as non-star--forming due to the
inability of the fiber to capture \ha\ emission in the outer disk of
the galaxy.

To study aperture bias in our SDSS galaxy sample, we have split the sample
discussed in Section \ref{selectioncriteria} into three subsamples as a
function of redshift: $0.05 \le z < 0.072$, $0.072 \le z < 0.085$, and $0.085
\le z < 0.095$. We chose these redshift shells because they represent equal
volumes. In Table \ref{aperture}, we show the median and $75^{th}$ percentile
of the \ha\ EW distribution for these three redshift shells as a function of
galaxy type.  First, we analysed all galaxies in each shell and find no
evidence for a systematic increase in the median and $75^{th}$ percentile with
redshift. Next, we restricted the analysis to just late--type galaxies using
the (inverse) concentration index because we expect such galaxies to have the
largest aperture bias. Once again, we find that the median and $75^{th}$
percentile of the \ha\ EW distribution of these galaxies does not
significantly increase with redshift. As a final check for aperture bias, we
further split the sample of late--type galaxies discussed above using their
Petrosian $g^*-r^*$ color and physical size (as defined by the Petrosian
radius that contains 90\% of the galaxy light in the $r^*$ band), and search
for the largest observed increase in the median and $75^{th}$ percentile of
the \ha\ EW distribution with redshift. The result of this search is given in
Table \ref{aperture} (labelled ``Worst Case'') where we find the smallest
($<8\ h^{-1}_{75}$), bluest ($g^*-r^*<0.75$), late--type galaxies possess the
largest systematic increase in the median and $75^{th}$ percentile of the \ha\
EW distribution with redshift. This is reasonable because for these
intrinsically small late--type galaxies, a larger fraction of the light
captured by the fiber is from the disk.

Thus the magnitude of any aperture bias is, at worst, 6\AA\ in the \ha\ EW and
only affects a small fraction of the late--type galaxies (see Table
\ref{aperture}).  We use a KS test to compare the distribution of local
(projected) densities observed in the three redshift shells for the subsample
of small, blue, late--type galaxies discussed above, and find no statistical
evidence for any difference in these distributions. This result indicates that
this subsample of galaxies inhabits the same environments at all redshifts and
that our conclusions are not significantly affected by any aperture bias.

We note here that the recent work of \citet{MW00} and \citet{Bart01} both
suggest that star--formation in cluster galaxies may be more centrally
concentrated than in field galaxies. If true, this could introduce an aperture
bias but in the wrong direction to explain our results, {\it i.e.},  it would
reduce the contrast between the cluster and field SFR distributions, because
we would be underestimating the SFR in field galaxies by a larger factor than
for cluster galaxies.  Therefore, the density--SFR trends we observe will only
be enhanced if this effect is important.

\subsection{Reddening}
Galactic extinction has been corrected for using the \citet{SFD} maps as
discussed in Stoughton et al. (2002), while no correction has been made for
possible intra--cluster extinction because this is expected to be small or
non--existent \citep{B+95,SKLD}.  More worrisome, however, is internal dust
extinction in each of the sample galaxies  and its effect on the
observed SFR \citep{Hopkins01,S+01,CL_ext}.  To address this possible bias, we
present in Figure \ref{density2} the median and 75$^{th}$ percentile of the
SFR as a function of density, for three different SFR prescriptions.  The
first is the ``standard'' derivation, based on an empirical methodology that
uses the \citet{K98} relation as given in Eqn 1., assuming one magnitude of
extinction for all galaxies.  The second, which is the prescription we have
adopted throughout this paper, is the same but corrected for SFR--dependent
reddening using the empirically derived relations of \citet{Hopkins01}.
Finally, we use the direct modeling approach of \citet{CL}, who provide a more
complete model for the dust, gas and ionizing radiation in galaxies based on a
combination of emission lines. However, the Charlot \& Longhetti prescription
requires that the \oii\ , \ha\ , H$_{\beta}$, [O{\sc iii}], \nii\ and [S{\sc ii}]
rest-frame EWs are larger than 5\AA, which demands high signal--to--noise
data and the presence of strong emission-lines. As a result, only 85 galaxies
satisfy the \citet{CL} prescription. For the remainder we just use the
corrected SFR of \citet{Hopkins01}. Therefore, in Figure \ref{density2}, the
corrected SFR and \citet{CL} SFR are very similar.

The fact that all three models show a trend of SFR with density suggests that
the dust properties of galaxies do not vary sufficiently with radius to give
rise to the observed trends.  Finally, in Figure \ref{densityew}, we show
that the \oii\, EW distribution as a function of density is similar to the
result obtained using \ha\ .  This is encouraging because the \ha\ and \oii\ 
lines have very different dust and metallicity sensitivities, come from
different star--forming regions in the galaxies, and are separated by nearly
3000\AA\ in wavelength.


\subsection{Flux Calibration}
As discussed in Stoughton et al. (2002), a 4--minute smear exposure is taken
for every SDSS spectroscopic science observation. This short exposure is
designed to obtain a more accurate spectrophotometric calibration as the
telescope is moved slightly during the smear exposure, effectively increasing
the nominal 3 arcsecond fiber aperture to a $5$ by $8$ arcsecond$^2$
aperture. This produces a low signal--to--noise spectrum of the galaxy, but
for a larger aperture, that then allows us to better correct the spectral
continuum radiation for the affects of seeing and atmospheric refraction.
Emission lines are excluded from the smear spectrum, so only the continuum of
the science exposure is effected by this procedure. This could introduce a
systematic bias into the EW measurements because the smear correction is only
changing the continuum shape relative to the unchanged emission line
properties.  Smear exposures are now a standard part of the SDSS observing
strategy.  However, for the EDR, only half of the data were taken with a
smearing exposure (see Table 3 in Stoughton et al. 2002) so we have a unique
opportunity to study the effect of this observing strategy on the line
measurements of SDSS galaxies. Using a Kolmogorov-Smirnov test, we were unable
to detect any statistically--significant difference between the two
distributions (smeared and non-smeared data) for \ha\ EW, \oii\ EW and the
SFR. This test suggests that the smear correction does not systematically bias
these measurements.

\subsection{Monte Carlo Simulations}
As a final test of our results, we performed 100 bootstrap simulations of our
data which involved randomly re--assigning (with replacement) the SFR of each
galaxy in our sample. The results of this test are consistent with no observed
correlation between SFR and local galaxy density.  Thus, the observed
correlation between SFR and density in our data is not an artifact of the
geometry or design of the sample because such systematic effects (minimum
separation of fibers, boundary effects) are still present in our bootstrap
simulations and would thus still appear as a correlation between SFR and
density if significant.


\clearpage
\begin{deluxetable}{ccccccc|l}
\tablenum{1} \tablewidth{0pt} \tablecaption{Groups
and clusters of galaxies used in this paper. 
\label{tabclus}
}
\tablehead{ 
\colhead{RA (J2000)} & 
\colhead{DEC (J2000)} & 
\colhead{z} & 
\colhead{$\sigma_r(1)$} & 
\colhead{$\sigma_r(2)$} & 
\colhead{N} &
\colhead{${\rm L_x(44)\,h_{75}^{-2}}$} &
\colhead{Name}\\
\colhead{(degrees)} & 
\colhead{(degrees)} & 
\colhead{} & 
\colhead{km s$^{-1}$}& 
\colhead{km s$^{-1}$}&
\colhead{} & 
\colhead{erg s$^{-1}$} &
\colhead{} \\
} 
\startdata 
5.853 & -0.857 & 0.063 & 225 & 430 & 17 & 0.02 &\\ 
7.304 & -0.199 & 0.059 & 345 & 515 & 24 & 0.16 &ZwCl 0027.0-0036 \\ 
15.574 & -0.379 & 0.050 & 312 & 246 & 7 & 0.07 &\\ 
20.039 & 0.032 & 0.077 & 223 & 351 & 15 & 0.04 &\\ 
22.852 & 0.627 & 0.079 & 504 & 761 & 39 & 0.39 &A208 (z=0.0798) \\ 
23.636 & -0.675 & 0.079 & 409 & 592 & 42 & 0.16&\\ 
24.424 & -0.440 & 0.056 & 298 & 556 & 18 &  & ZwCl 0134.8-0045\\ 
30.990 & 0.270 & 0.076 & 228 & 254 & 10 & 0.04 &A299\\
$^*$151.942 & 0.496 & 0.096 & 542 & 563 & 64 &1.23 &A933 (0.0956) \\ 
191.831 & 0.143 & 0.088 &1243 & 1450 & 236 & 1.96 & \\
197.967 & -0.751 & 0.084 & 878 & 1003 & 85 & 3.75 &\\ 
199.787 & -0.873 & 0.083 & 677 & 771 & 67 & & ZwCl 1316.4-0044\\ 
201.575 & 0.232 & 0.082 & 468 & 528 & 36 & 1.07 & RX J1326.3+0013 \\
        &       &       &     &     &    &      & (z=0.0820)\\ 
206.317 & 0.098 & 0.088 & 345 & 330 & 32 &  & ZwCl 1342.4+0016\\ 
217.565 & 0.337 & 0.055 & 297 & 354 & 14 &  & \\ 
227.302 & -0.250 & 0.090 & 682 & 701 & 84 & 0.34 & A2026 (z=0.0876)\\ 
227.734 & -0.103 & 0.091 & 567 & 573 & 57 &  & A2030 (z=0.0919)\\ 
\enddata \\
\footnotesize{We have included in our analysis clusters marked with
a $^*$ even though their mean redshift is beyond the nominal redshift limit of
the sample defined in the text. This is because some of their cluster galaxies
are in the sample limits.}

\end{deluxetable}

\begin{deluxetable}{ccccccc|l}
\tablenum{2} \tablewidth{0pt} \tablecaption{Median and $75^{th}$ Percentile of the \ha\ EW Distribution as a Function of Redshift and Galaxy Type
\label{aperture}
}
\tablehead{ 
\colhead{} & 
\colhead{$0.05 < z < 0.072$} & 
\colhead{$0.072 < z < 0.085$} & 
\colhead{$0.085 < z < 0.095$} \\
\colhead{} &
\colhead{Median, $75^{th}$ (Numb.)} & 
\colhead{Median, $75^{th}$ (Numb.)} & 
\colhead{Median, $75^{th}$ (Numb.)} \\
}
\startdata 
All Galaxies & 8.1, 12.3 (1632) & 7.9, 12.8 (2849) & 8.4, 13.1 (1977) \\
Late--type   & 15.3, 20.4 (632) & 16.1, 23.3 (1045) & 16.0, 23.1 (809) \\
Worst Case  &  27.1, 35.8 (130) & 28.5, 36.1 (205)  & 31.5, 41.8 (133) \\  
\enddata \\

\end{deluxetable}

\begin{deluxetable}{ccc}
\tablenum{3} \tablewidth{0pt} \tablecaption{The Mean and Median Percentage Difference in SFR for Duplicate Observations as a function of SFR.
\label{errors}
}
\tablehead{ 
\colhead{SFR (\Msun$\, {\rm yr^{-1}}$)} & 
\colhead{Median} & 
\colhead{Mean} 
}
\startdata 
    $<0.05$  &     25\%    &   141\% \\
     0.05--0.15  &    36\%   &    64\% \\
     0.15--0.90  &    16\%   &    29\% \\
      $>3.0$  &    18\%    &   19\% \\
\enddata \\

\end{deluxetable}


%
%

\begin{figure}
\epsscale{1.05}
\plottwo{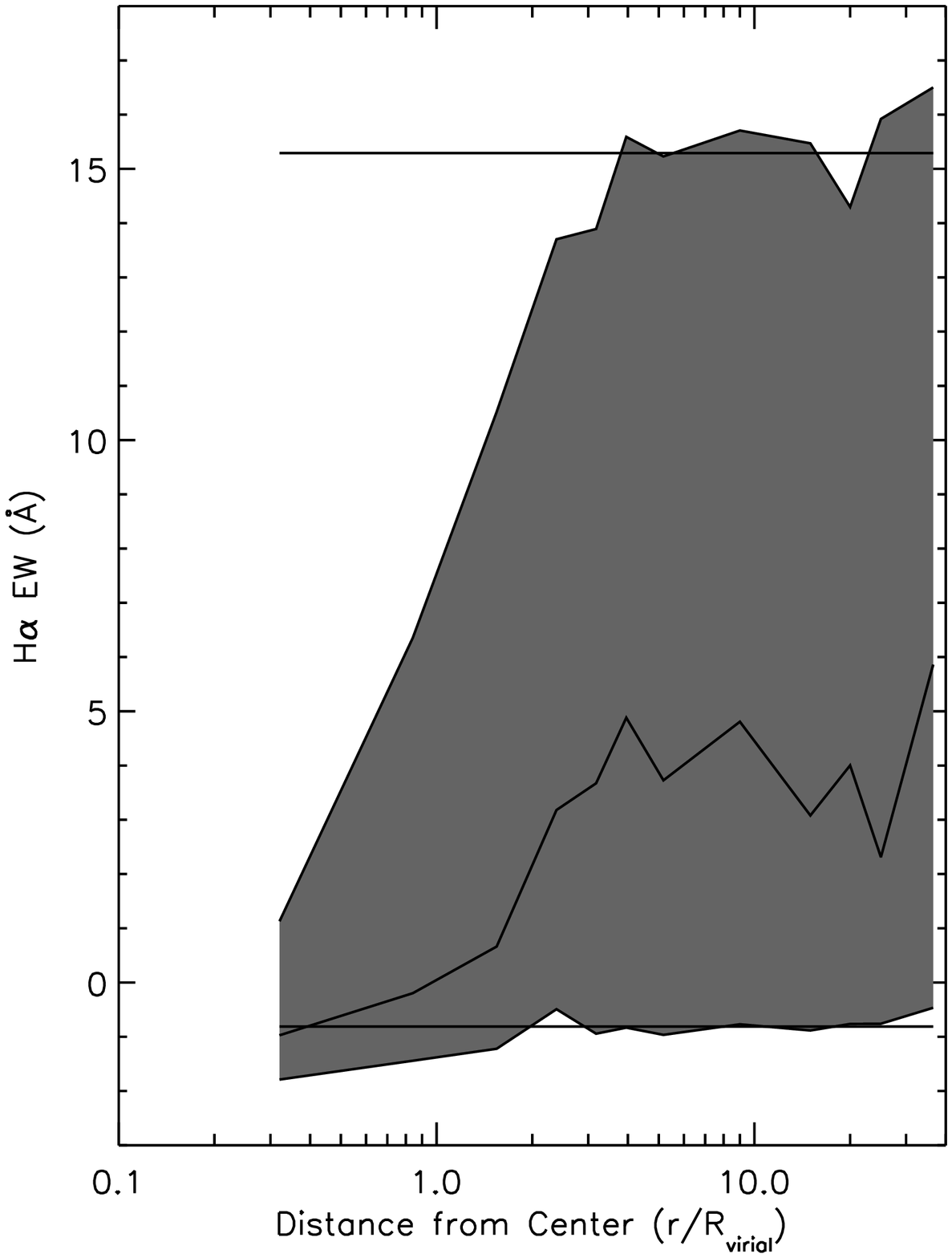}{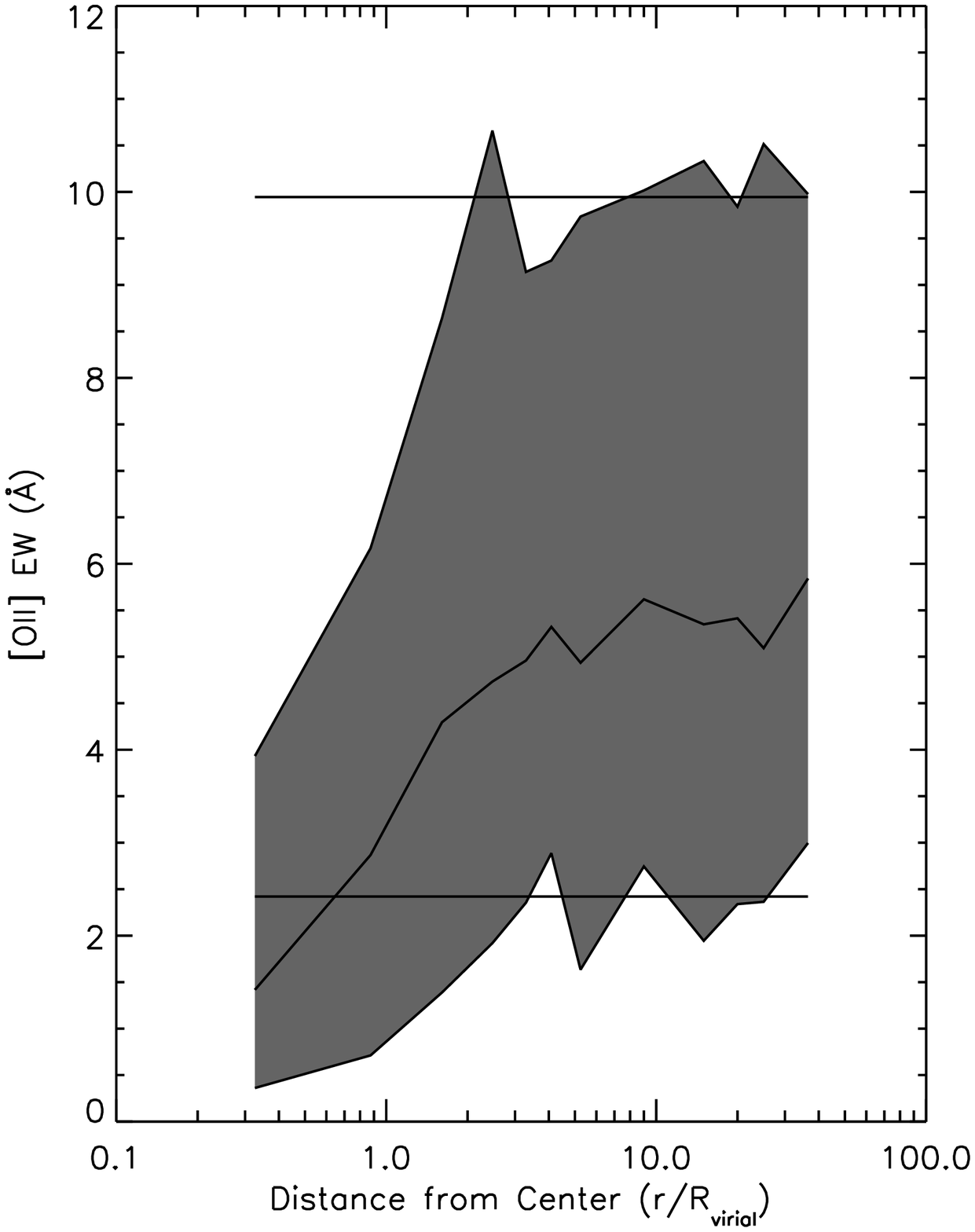}
\caption{(Left) The shaded area is the distribution of \ha\ EW as a function of
(projected) clustercentric radius. We have scaled the projected radial
distances by the virial radii for the appropriate cluster $\sigma_r(1)$ (see
Table \ref{tabclus}). The top and bottom of the shaded area represents the
75\,$^{th}$ and 25\,$^{th}$ percentile of the EW distribution, while the light
grey line inside the shaded area is the median of the distribution. The
straight lines at the top and bottom of the shaded area show the 75\,$^{th}$
and 25\,$^{th}$ percentile of the EW distribution for the field population
respectively (see text for definition of the field population).  (Right) The
same, but for the \oii\ EW distributions. In each plot, there are 180 galaxies
per bin.
\label{clusterew}  
}
\end{figure}

\begin{figure}
\epsscale{1.05}
\plottwo{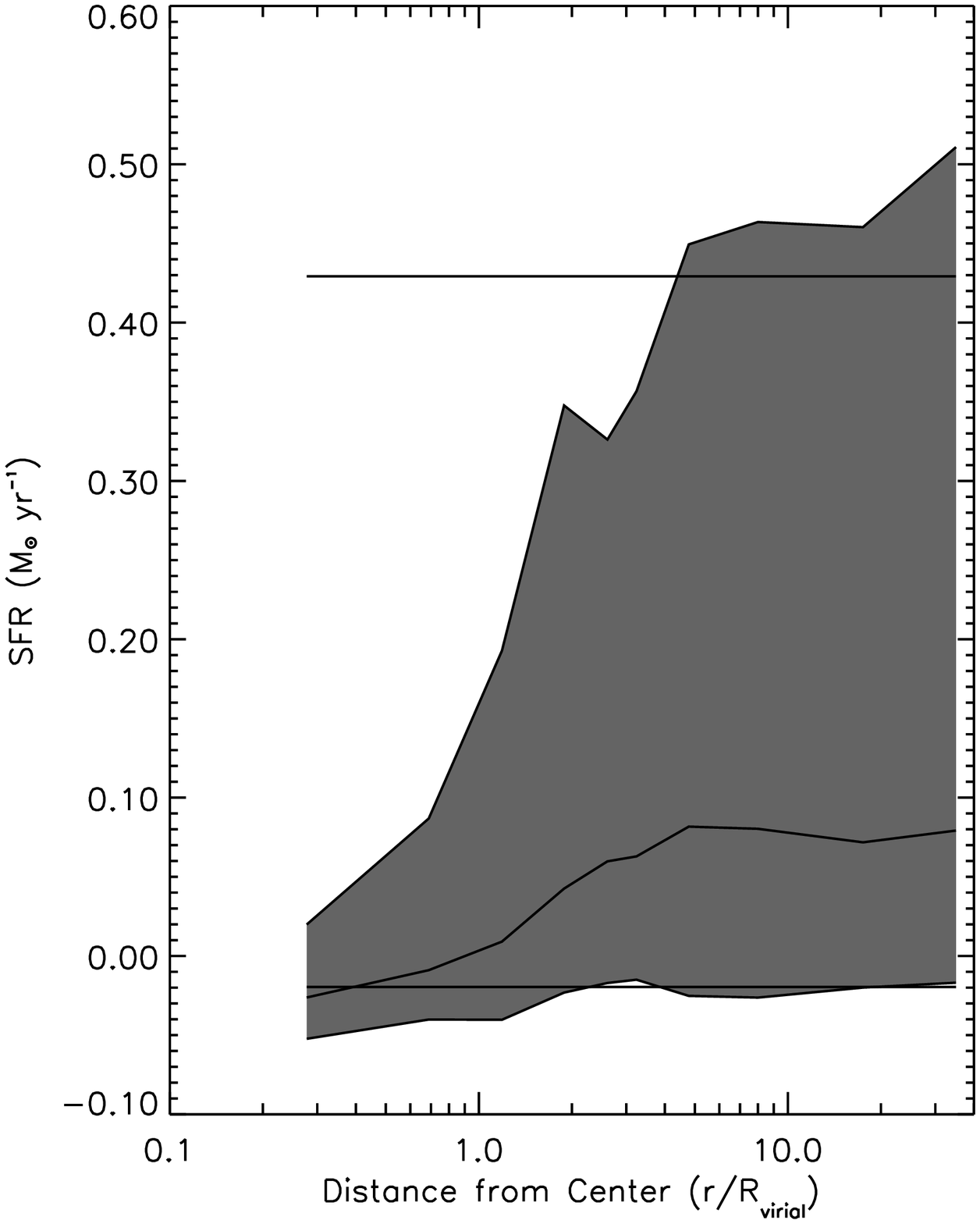}{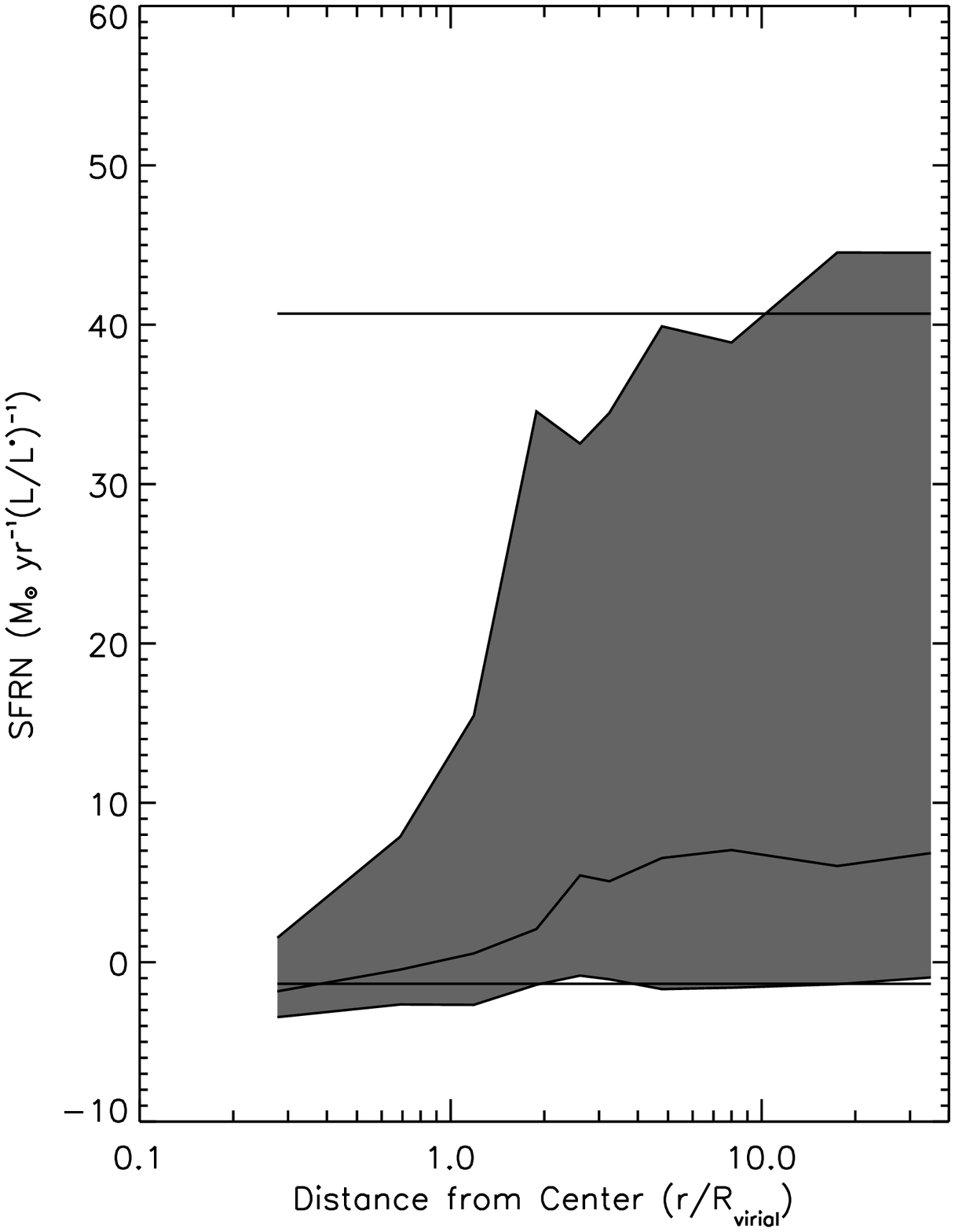}
\caption{The same as Figure~\ref{clusterew}, but for the
corrected SFR (left) and the normalized,
SFRN (right) distributions.  Each bin contains 150 galaxies.
\label{clustersfr}  
}
\end{figure}

\begin{figure}
\epsscale{0.8}
\plotone{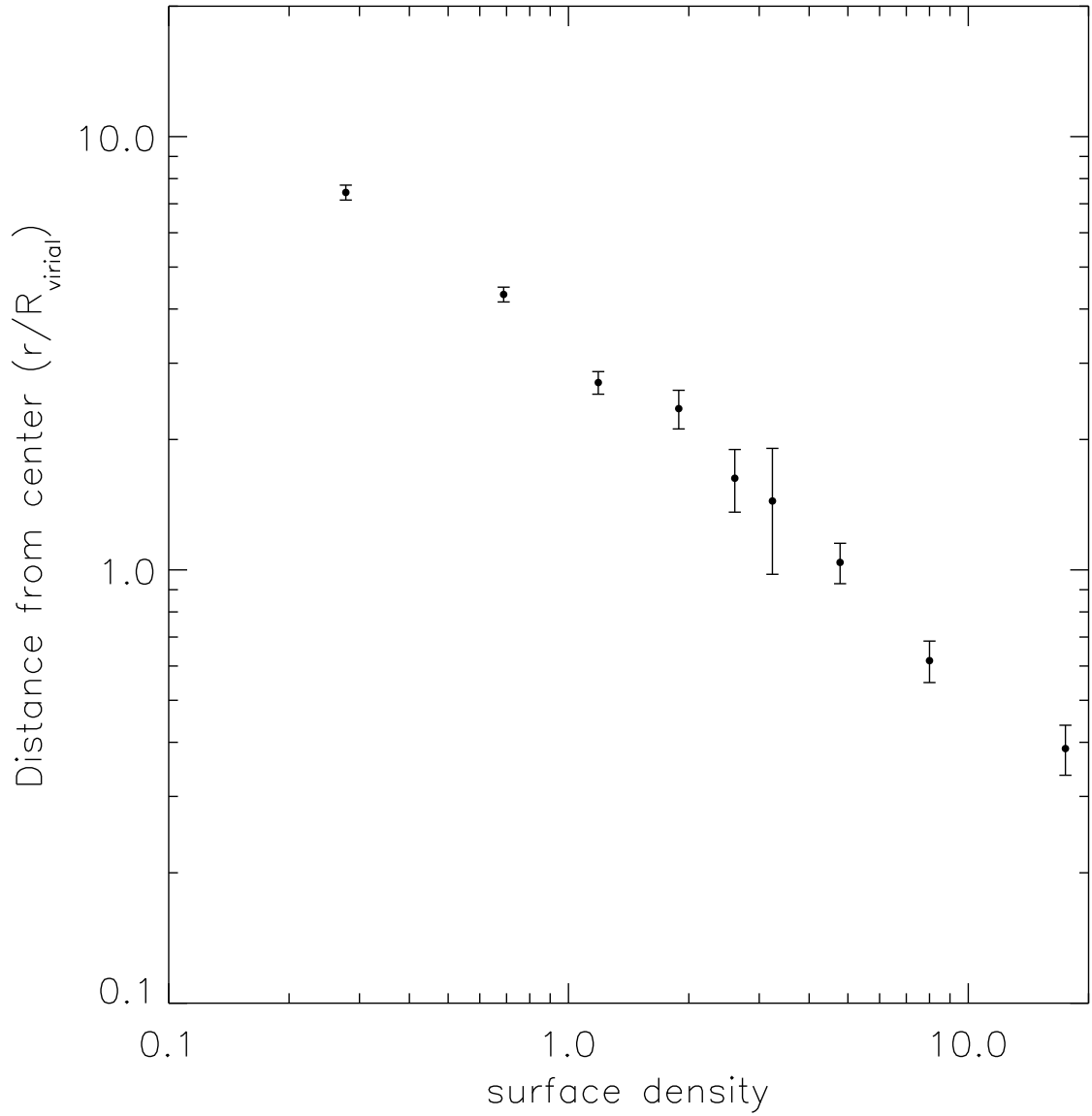}
\caption{The clustercentric distance as a function of the local galaxy density
for galaxies in our cluster sample. The points are the median of each bin with
the error on the median from bootstrap re--sampling.  \label{map1} }
\end{figure}

\begin{figure}
\epsscale{1.2}
\plottwo{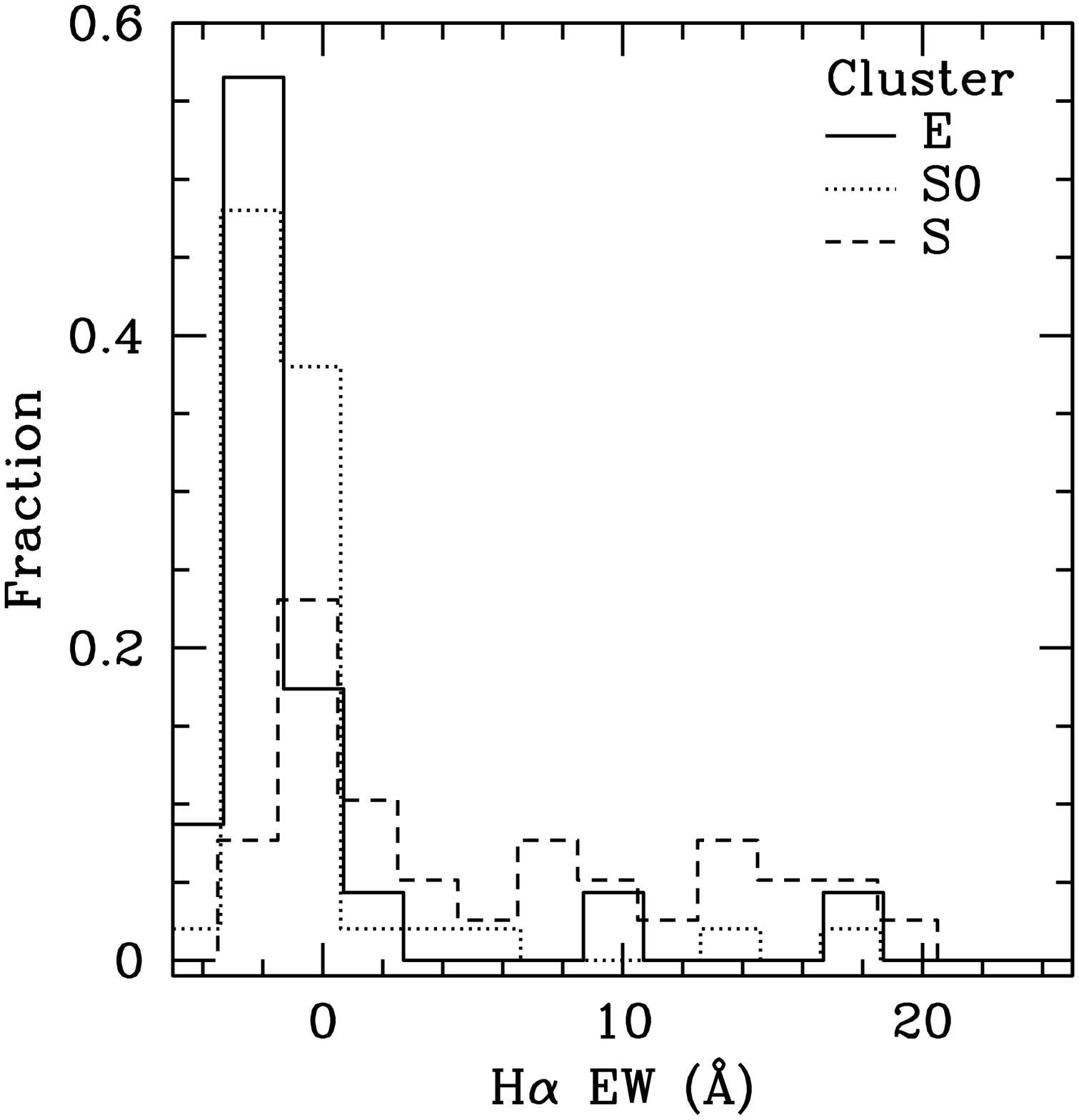}{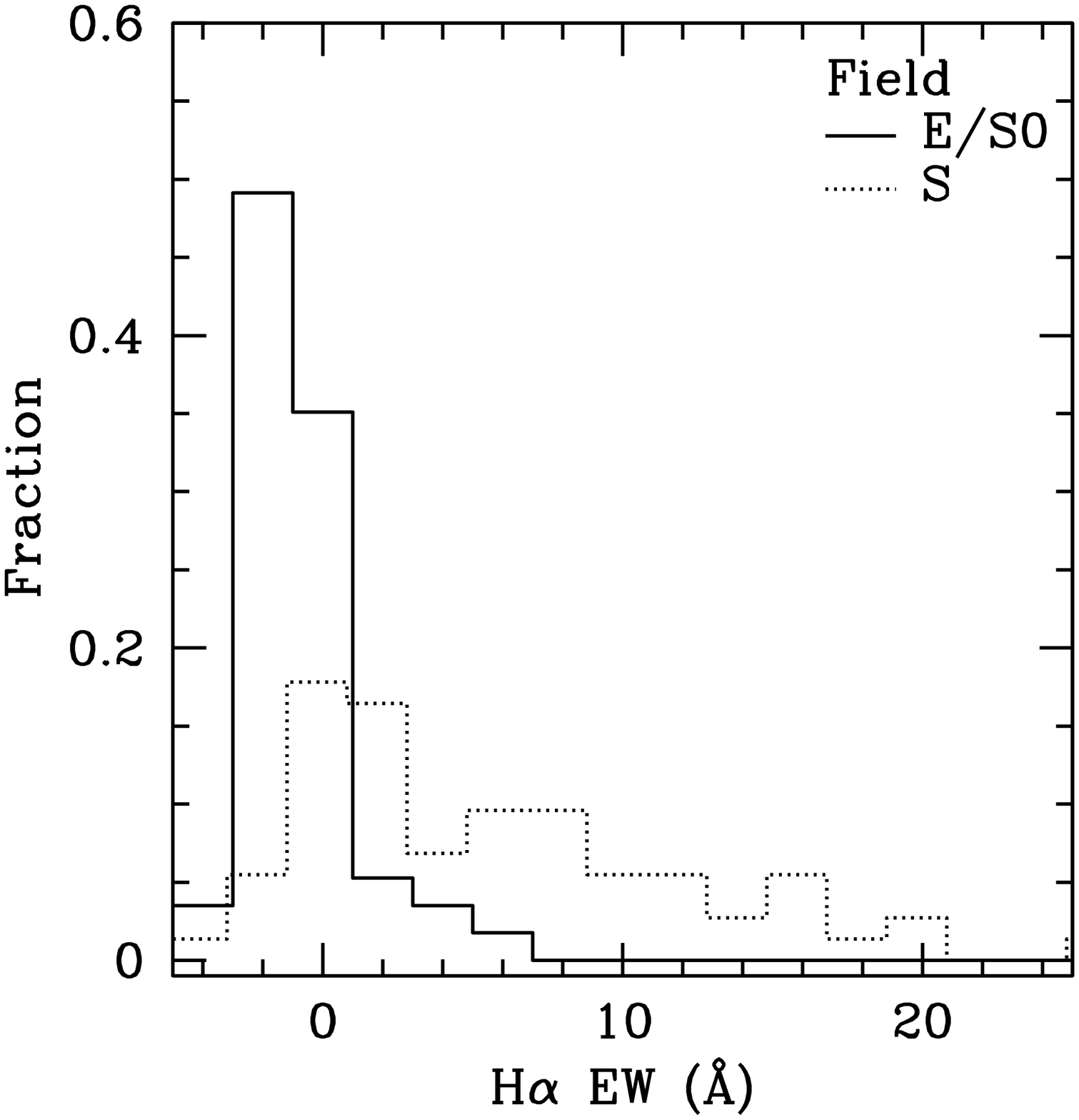}
\caption{(Top) We show the observed distribution of \ha\ EWs as a function of
morphological classification for 114 cluster galaxies serendipitously observed
by the SDSS. The morphologies come from Dressler (1980) and include
ellipticals (E), lenticulars (S0) and spirals (S). The \ha\ EWs come from the
SDSS database. (Bottom) The distribution of \ha\ EWs for the Shimasaku et
al. field galaxy sample. We only show late (S) and early--type (E/S0) galaxies
(see text).
\label{morphs} 
}
\end{figure}

\begin{figure}
\epsscale{0.65}
\plotone{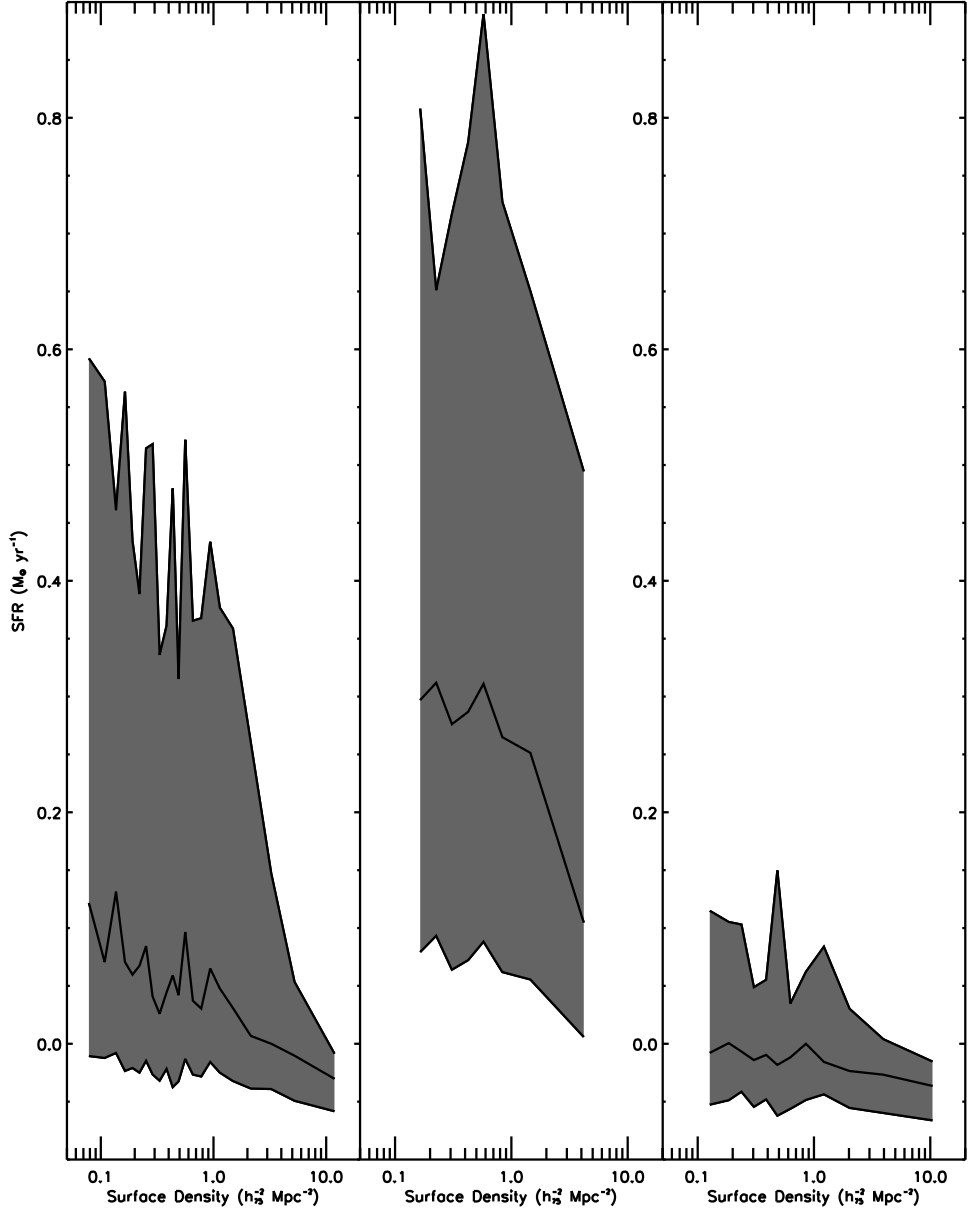}
\caption{(Left panel) The same as shown in Figure \ref{densitysfr}.  (Middle
panel) The density--SFR for late--type galaxies ({\it i.e.}, with $C>0.4$ as
discussed in Section \ref{densityanalysis}). As in other plots, the top of the shaded area is the 75\,$^{th}$ percentile of the distribution, while the bottom of the shaded region is the 25\,$^{th}$ percentile of the distribution. The median is shown as a black line near the middle of the shaded region. (Right Panel) The density--SFR for early--type galaxies ({\it i.e.}, with $C\le0.4$).
\label{newfigure}
}
\end{figure}
\begin{figure}
\epsscale{0.8}
\plotone{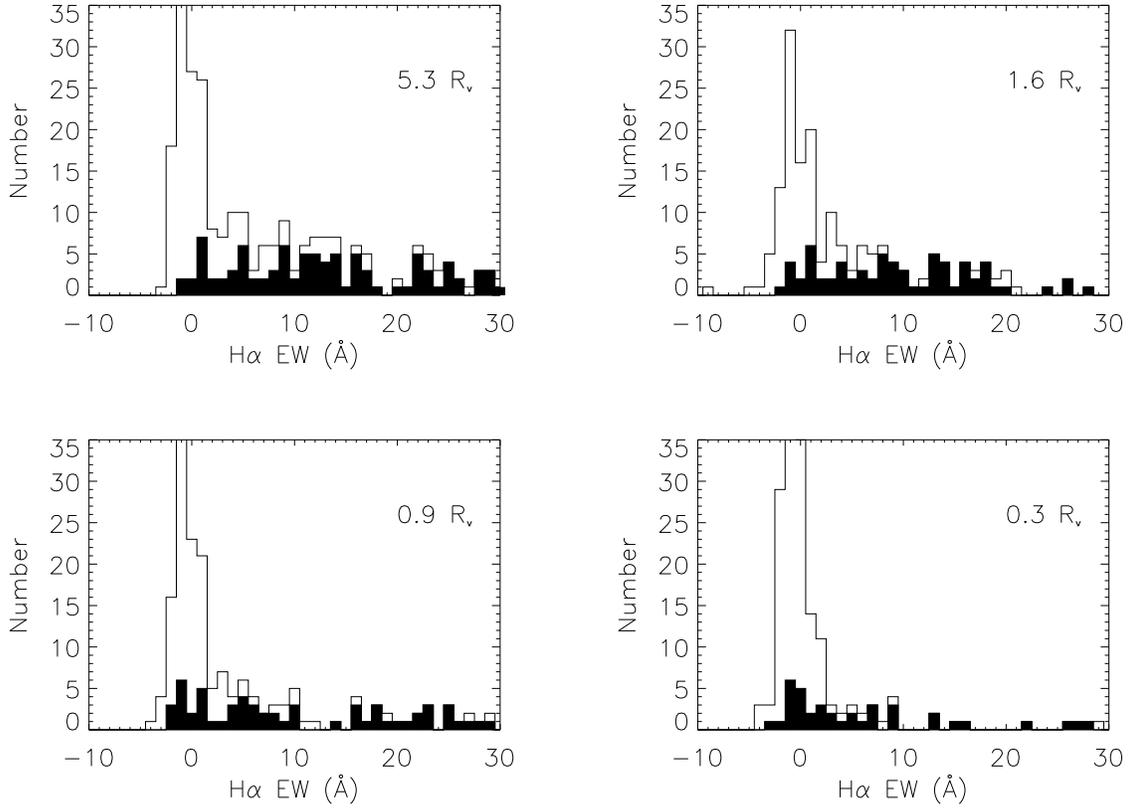}
\caption{The distribution of \ha\ EWs as a function of clustercentric radius
and galaxy (inverse) concentration index.  The unfilled histograms represent
the whole distribution of \ha\ EWs regardless of morphological classification.
The filled histograms are the distribution of \ha\ EWs for galaxies with an
(inverse) concentration index of $C>0.4$ \citep{SDSS_morph} and are thus
classified as late--type galaxies.
\label{histo}
}
\end{figure}

\begin{figure}
\epsscale{0.8}
\plotone{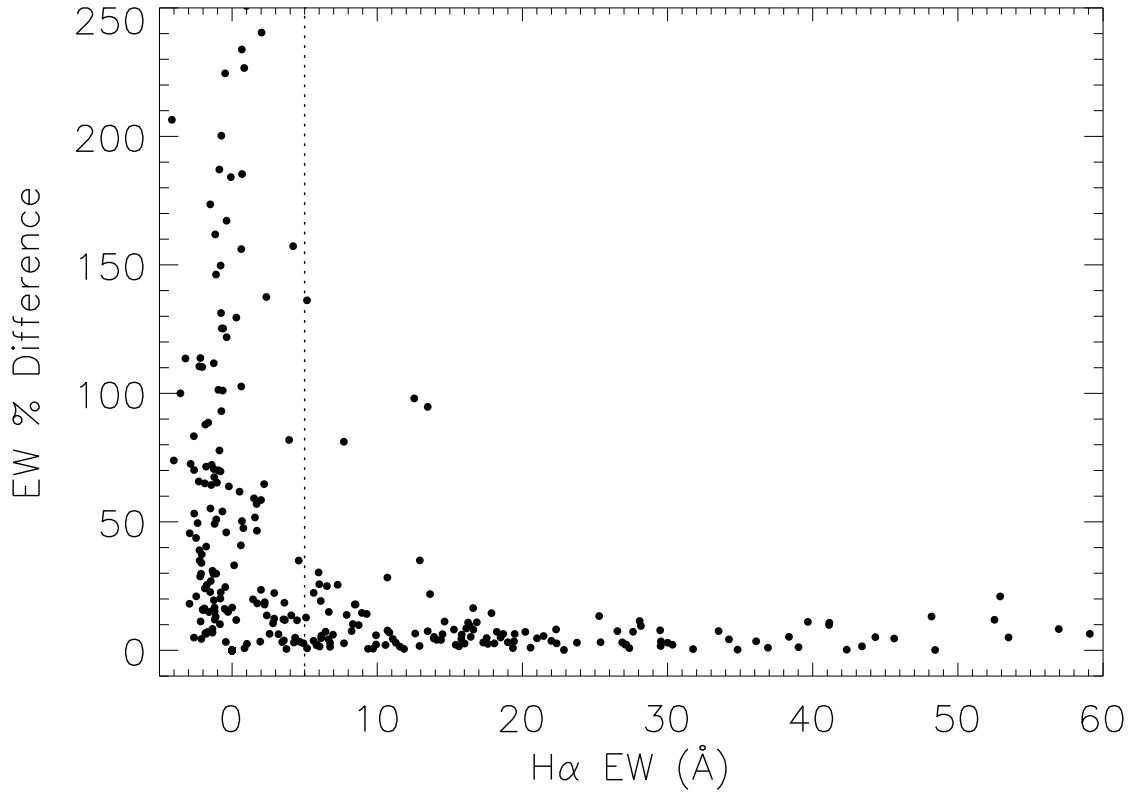}
\caption{We show here the percentage absolute difference for the EW of \ha\,
as function of the observed \ha\ EW, for EDR galaxies that were observed
twice in the SDSS and analysed independently through the official SDSS
SPECTRO1D analysis pipeline. The dashed vertical line is at \ha=5\AA, where the
errors begin to stabilize below 20\%.
\label{bernardi}
}
\end{figure}

\begin{figure}
\epsscale{0.8}
\plotone{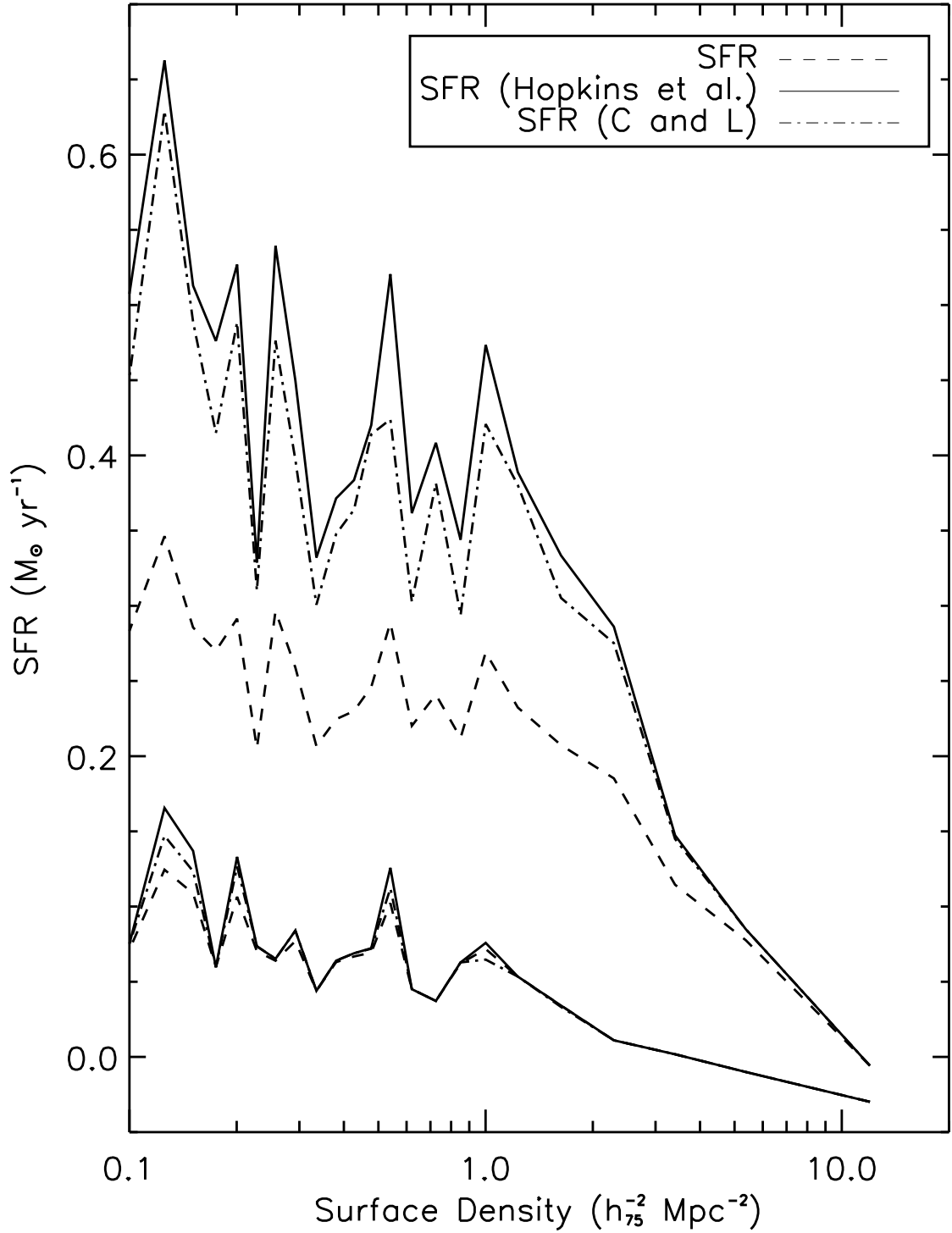}
\caption{The SFR of galaxies as a function of density for three different
extinction corrections.  The dashed lines are for the median and 75\,$^{th}$
percentile for the standard \citet{K98} formalism, assuming 1 magnitude of
extinction.  The solid lines are the median and 75\,$^{th}$ percentile for the
SFR-dependent dust correction using \citet{Hopkins01}. For the median, these
two SFR indicators are very similar because the empirical corrections of
\citet{Hopkins01} are only significant for high SFR galaxies as shown by their
difference in the 75$^{th}$ percentile. The dashed-dot line shows the results of
the \citet{CL} model that uses information from six emission lines to
constrain the dust properties of the galaxies.
\label{density2}
}
\end{figure}

\end{document}